# Lead-Time Simple Exponential Smoothing as a Configurable, Uncertainty-Aware Framework for Adaptive Radiotherapy Monitoring

**Robert Boyd[1,2] and Wolfgang A. Tomé[1,2]**

[1]Montefiore Einstein, 111 East 210$^{\text{th}}$ Street, Bronx, NY 10467
[2]Institute for Onco-Physics, Albert Einstein College of Medicine, 1300 Morris Park Ave, Bronx, NY 10461

**Declarations of interest:** none

Short Title: **Uncertainty-Aware Lead-Time SES for Adaptive Monitoring**

**Address of Correspondence:**

Professor Wolfgang A. Tomé, Ph.D., FAAPM, FASTRO
Montefiore Einstein
Institute for Onco-Physics
Albert Einstein College of Medicine
Block Building Room 106
1300 Morris Park Ave
Bronx, NY 10461
Email: wolfgang.tome@einsteinmed.edu

August 25, 2026

# Abstract

**Background:** Adaptive radiotherapy (ART) generates sequential temporally ordered anatomical and dosimetric data throughout the course of treatment. Standard dose-difference summaries describe observed deviations but do not explicitly represent uncertainty in fractions that remain unobserved. Quantifying this uncertainty can provide additional information for adaptive monitoring.

**Purpose:** To introduce lead-time simple exponential smoothing (LT-SES) as a configurable, uncertainty-aware analysis framework for ART monitoring in which a prospectively specified smoothing parameter $\alpha$ controls how strongly new observations update the current level, allowing the temporal response to be selected according to the clinical need for rapid response and the degree of confidence that the current level will persist.

**Methods:** Using the standard innovations formulation of simple exponential smoothing (SES), closed-form expressions were derived for the predictive mean and variance of the remaining-course aggregate, allowing prediction intervals and conditional threshold probabilities for the final-course metric to be updated after each fraction. Each dose-metric time series was evaluated at prespecified $\alpha$-values from 0.1 to 1.0, and results for each candidate setting were summarized across time series; $\alpha$ was not fitted separately to individual time series in the primary interval analysis. Empirical coverage of nominal 95% and 99% intervals was characterized in 340 organ-at-risk mean-dose time series from 32 prostate and 23 head-and-neck treatment courses across remaining-course horizons of $5, 10, 15$, and 20 fractions using a running-mean benchmark. Two parotid time series were evaluated using $\alpha = 0.25, 0.50$, and 1.0 to illustrate threshold probabilities, probability maps, and probability-weighted rank scores under alternative temporal-response settings.

**Results:** Prediction intervals widened with forecast horizon and, among LT-SES configurations, with increasing values for $\alpha$. Undercoverage was greatest for the running-mean benchmark and the lowest positive $\alpha$ values and became more pronounced at longer horizons, whereas $\alpha$-values $\geq 0.4$ produced wider intervals with coverage generally above nominal. These findings characterize the behavior of the evaluated configurations in this cohort. The parotid examples considered demonstrate how different $\alpha$-values affected the model response to short-term variation and a sustained level shift, with lower $\alpha$-values providing greater stability and higher $\alpha$-values responding more rapidly to recent changes. In the two illustrative parotid time series, probability weighting produced greater numerical separation than the conventional scalar comparator and assigned the sustained-shift time series a higher score several fractions before the scalar ratio crossed unity, with the same qualitative ordering across the tested $\alpha$-values.

**Conclusion:** LT-SES places a projected scalar threshold difference in the context of the time series-specific interfractional innovation scale and the uncertainty associated with the fractions that remain unobserved. The smoothing parameter $\alpha$ provides a prospectively configurable temporal response: larger values favor rapid response to recent observations, whereas smaller values place greater persistence on the current level estimate. Selection of $\alpha$ should reflect the clinical monitoring objective and the degree of confidence that the current level will persist. The resulting probability-scale outputs and rank scores add a relative, uncertainty-aware layer to conventional scalar dose-metric monitoring.

# 1 Introduction

Adaptive radiation therapy was formalized as a closed-loop process in which data acquired during treatment are feed back into treatment management [1]. Subsequent work showed that systematic anatomical changes, including tumor regression, parotid shrinkage, and treatment-related weight loss, can alter target and organ-at-risk (OAR) dose distributions during a course of fractionated radiation therapy [2, 3]. Modern ART practices now range from scheduled offline review and replanning to daily online adaptation, with substantial variation in imaging, dose-recalculation, dose-accumulation, and decision protocols [4, 5].

The ART process can be separated into three interconnected layers (see Figure 1). The measurement layer produces daily anatomical, structural, and dosimetric data. The analysis layer converts those data into summaries, trends, projections, and estimates of uncertainty. The decision-making layer combines the analysis with clinical goals, persistence rules, institutional resources, and professional judgment to determine whether treatment should continue, receive enhanced review, or be adapted. Published ART monitoring methods have included trend analysis, statistical process control, projected course metrics, and explicit uncertainty bounds for deformable accumulation [5–10]. Nevertheless, to our knowledge, the temporal uncertainty associated specifically with the unobserved yet to be delivered fractions is not routinely reported as an output of published ART analysis workflows. This temporal predictive uncertainty is conceptually distinct from Monte Carlo dose-calculation uncertainty, robust-optimization scenarios, contouring uncertainty, or deformable-registration uncertainty.

Simple exponential smoothing (SES) is a transparent local-level forecasting model whose recursive level update forms an exponentially weighted moving average of prior observations. It has been used in medical physics primarily through statistical process control [11, 12]. In forecasting and inventory applications, lead-time methods extend the one-step forecast to the aggregate quantity expected over a future time interval [13–16]. The analogous problem in a fractionated treatment course of radiation therapy is to estimate the aggregate of a scalar dose metric over the remaining, unobserved yet to be delivered fractions and to quantify the uncertainty in that estimate.

The purpose of this work is to introduce lead-time SES, incorporate it into the adaptive monitoring process of radiation therapy, and to provide examples of its use in it. The contribution is intentionally limited to the analysis layer. We derive the predictive mean and variance of a remaining-course scaler dose metric under a Gaussian innovation model, characterize the empirical behavior of the resulting intervals in an available retrospective dataset, and LT-SES was evaluated at $\alpha = 0.25, 0.50$, and 1.0 to illustrate a range of temporal-response settings, from greater stability against short-term variation to greater responsiveness to recent observations.. The study was not designed to define a replanning threshold, establish overall plan superiority, incorporate all measurement and dose-accumulation uncertainties, or demonstrate clinical benefit.

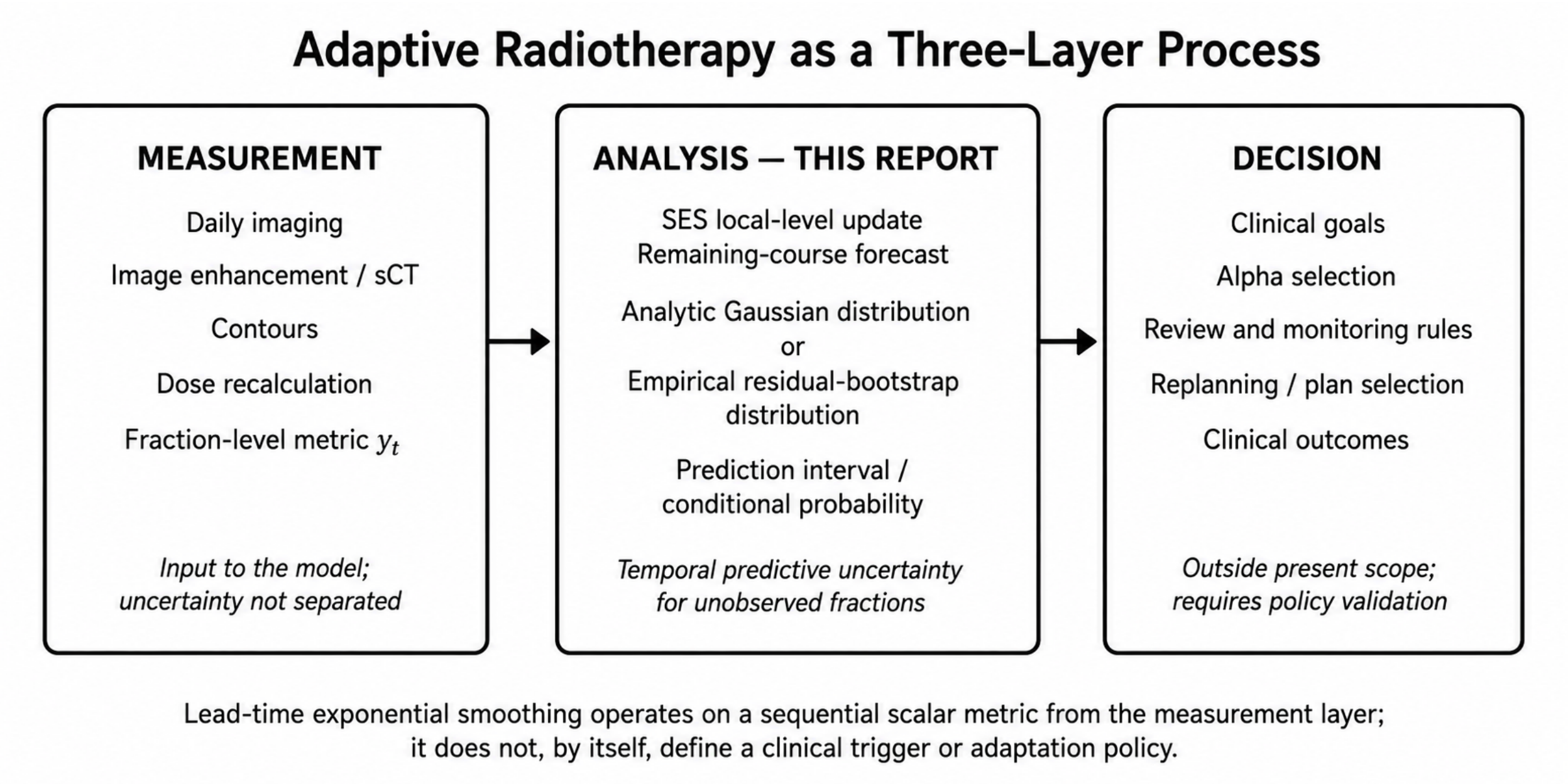

Figure 1: Adaptive radiotherapy represented as measurement, analysis, and decision-making layers. The present work is confined to the analysis layer. The smoothing parameter $\alpha$ controls how strongly new observations update the current level and may be selected for a monitored metric according to the clinical monitoring objective. The level and innovation variance are estimated from the individual metric time series. The sequential scalar metric is supplied by the measurement layer; clinical thresholds and intervention policies belong to the decision layer. In the figure, sCT denotes synthetic computed tomography.

# 2 Methods

## 2.1 Scope and Interpretation of the Scalar Dose Metric

Let $y_t$ denote the observed value of a prespecified scalar dose metric at fraction $t$. Examples of these include a fraction-specific structure mean dose, a maximum-dose surrogate, or a dose-volume histogram (DVH) ordinate evaluated on the daily anatomy. The measurement pipeline that produces $y_t$ may include daily imaging, image enhancement or synthetic computed tomography, contour propagation or editing, and dose recalculation. The forecasting model begins after that pipeline and treats the resulting time series as its input. For the purpose of this work clinical-goal evaluation is categorized into the following three tiers of increasing safety and monitoring complexity for patient treatment:

**Tier 1:** **Initial-plan metrics.** Metrics calculated from the approved treatment plan on the planning image. These metrics remain the most directly applicable to published clinical constraints because most clinical trials and dose–outcome relationships were based on the planned dose distribution. However, they do not account for anatomical or dosimetric changes during treatment.

**Tier 2:** **Fraction-averaged daily-anatomy metrics.** Metrics recalculated on each daily anatomy and summarized across fractions. This approach incorporates fraction-to-fraction anatomical and dosimetric variation without requiring deformable dose mapping. However, it does not preserve the spatial history of dose deposition, and averaged nonlinear DVH metrics are not necessarily equivalent to metrics derived from accumulated dose. It can, however, provide an additional safety-monitoring layer beyond reliance on tier 1 alone.

**Tier 3:** **Reference-volume accumulated-dose metrics.** Fractional dose distributions are mapped to a common anatomy and then used to recalculate the final metric. This approach provides the most spatially comprehensive estimate of delivered dose.

The retrospective analyses in this work principally concern tier 2. For an additive quantity, the fraction-level values may be summed directly. For a nonadditive quantity such as $V_D$, a mean or aggregate of daily-anatomy values is a longitudinal analysis metric and is not mathematically equivalent to calculating $V_D$ from a deformably accumulated three-dimensional dose distribution.

## 2.2 LT-SES Cumulative Dose Metric Forecasting

As defined above let $y_s$ be the observed per-fraction dose metric delivered at treatment fraction $s = 1, 2, \cdots$. At any time $t$, the information set $\mathcal{F}_t = \{y_1, \ldots, y_t\}$ contains all metrics delivered up to and including fraction $t$. For a delivery horizon of $h$ additional fractions, we define the aggregate of the *remaining-course* (lead-time) dose metric of unknown future values $y_{t+j}$ as the random variable

$$R_{t+h} = \sum_{j=1}^{h} y_{t+j}. \tag{1}$$

We denote point forecasts of future scalar dose metrics by a hat. In particular, $\widehat{R}_{t+h|t}$ is the point forecast of $y_{t+h}$ based on observed data $\mathcal{F}_t$, and

$$\widehat{R}_{t+h|t} = E\left(R_{t+h} \mid \mathcal{F}_t\right) \tag{2}$$

is the corresponding forecast of the aggregate of the remaining-course dose metric. Throughout, $E$ and $V$ denote the conditional expectation and the variance conditioned on the observed information set $\mathcal{F}_t$. We use the simple linear innovations state space model

$$\begin{aligned} y_t &= \ell_{t-1} + \varepsilon_t \\ \ell_t &= \ell_{t-1} + \alpha\varepsilon_t \end{aligned} \tag{3}$$

where $\ell_t$ is the level at time t, $0 < \alpha \leq 1$ is the smoothing parameter, and $\varepsilon_t$ denotes the one-step innovation (or forecast error). As shown in Supplementary Section S1.1 in the ARIMA(0,1,1) framework, SES is the optimal linear one-step ahead forecast of the scalar dose metric per fraction process $\{y_t\}$. Note that we are estimating aggregate dosimetric quantities, such as the total accumulated mean dose, the total accumulated maximum dose to an organ at risk, the volume receiving a total dose D, $V_D$, or the total expected dose a PTV or CTV receives. Since forming these quantities involves summation of the per fraction dose metrics $y_t$, one knows from the central limit theorem that their cumulative distribution function ($CDF$) will tend towards the General Normal Cumulative Distribution function $\Phi\left(\frac{z-\mu}{\sigma}\right)$ as the prediction horizon increases. However, for small to moderate prediction horizons, one needs to estimate the $CDF$. It is shown in Supplementary Section S2 how this can be done using the measured one step innovations. In order to obtain analytic closed form expressions, it is assumed without loss of generality for the remainder of this section that the one-step innovations $\varepsilon_t$ are independent and identically distributed, following a normal distribution with mean zero and variance $\sigma^2$.

The conditional mean of the aggregate of a scalar dose metric at time $t + h$ is the sum of its point forecasts:

$$\widehat{R}_{t+h|t} = E\left(R_{t+h} \mid \mathcal{F}_t\right) = \sum_{j=1}^{h} \hat{y}_{t+j|t}. \tag{4}$$

For the simple linear innovations state space model, we are considering here this sum reduces to the following:

$$\widehat{R}_{t+h|t} = h\ell_t. \tag{5}$$

Each future innovation $\varepsilon_{t+h-j}$ contributes once directly to $y_{t+h-j}$, and to each of the next $j$ future fractions through the level update

$$\ell_{t+h-j} = \ell_{t+h-j-1} + \alpha\varepsilon_{t+h-j} \tag{6}$$

so, the same innovation is propagated into $y_{t+h-j+1}, \dots, y_{t+h}$, each time with coefficient $\alpha$. Thus, $\varepsilon_{t+h-j}$ has total weight $1 + j\alpha$ in the sum of the remaining-course (1) which yields the factor $(1 + j\alpha)^2$ in the conditional variance. Therefore, the conditional variance of the aggregate of the scalar dose metric is

$$V(R_{t+h} \mid \mathcal{F}_t) = \sigma^2 \sum_{j=0}^{h-1} (1 + j\alpha)^2. \tag{7}$$

Summing over $j$ yields

$$V(R_{t+h} \mid \mathcal{F}_t) = \sigma^2 h \left[ 1 + \alpha(h-1) + \frac{\alpha^2}{6}(h-1)(2h-1) \right]. \tag{8}$$

The interested reader is referred to the Supplementary Section S1.2 for details of these calculations. For notational convenience, let us define the following function of $h$ and $\alpha$:

$$\Omega(h, \alpha) \equiv \; h \left[ 1 + \alpha(h-1) + \frac{\alpha^2}{6}(h-1)(2h-1) \right], \tag{9}$$

then the conditional variance simply becomes

$$V(R_{t+h} \mid \mathcal{F}_t) = \sigma^2 \Omega(h, \alpha). \tag{10}$$

### 2.3 Initial Level

At each forecast origin t, the initial level $\ell_0$ was treated as an unknown initial state and re-estimated conditional on the smoothing parameter $\alpha$, using only the observations available through $t$. Under the additive Gaussian-innovation form of SES, maximizing the conditional likelihood is equivalent to minimizing the sum of squared one-step-ahead prediction errors [13]:

$$\ell_0^*(t, \alpha) = \max_{\ell_0} \mathcal{L}(\ell_0 \mid \alpha) = \min_{\ell_0} \sum_{s=1}^{t} \left| y_s - \ell_{s-1}(\ell_0, \alpha) \right|^2. \tag{11}$$

Here, $\ell_{s-1}(\ell_0, \alpha)$ is obtained by recursively applying equation (3) from the candidate initial level $\ell_0$. Because the SES state is linear in $\ell_0$, the optimization has the following closed-form solution (cf. Ref [17]):

$$\widehat{\ell}_0(t, \alpha) = \frac{\sum_{k=1}^{t} w_k \; [y_k - c_k(\alpha)]}{\sum_{k=1}^{t} w_k}, \quad w_k \equiv (1-\alpha)^{k-1}, \quad c_k(\alpha) \equiv \alpha \sum_{j=1}^{k-1} w_{k-j-1} y_j. \tag{12}$$

Note that regardless of the value of $\alpha$, $c_1(\alpha) = 0$ (since the sum is empty) and $w_1 = 1$. For each forecast origin t and each value of $\alpha$, $\widehat{\ell}_0(t, \alpha)$ was recalculated, after which equation (3) was reapplied through observation $t$ to obtain the current level and the one-step-ahead innovations used to construct the predictive distribution. Lastly, let us look at the boundary cases of $\alpha = 1$ and $\alpha = 0$. For $\alpha = 1$ equation (12) becomes undefined and in this case one simply sets without loss of generality $\widehat{\ell}_0(t, 1) \equiv y_1$, which is the result when evaluating equation (12) for $t = 1$ and $\alpha < 1$. Note while $\alpha$ needs to be strictly larger than zero for our LT-SES model to be well defined, it is however of interest for what follows to look at what happens for the case if one assumes that $\alpha = 0$, in that case a moments reflection shows that $\widehat{\ell}_0$ simply becomes the mean of the observations available through $t$.

### 2.4 Prediction Interval

To make it clear that the random variable $R_{t+h}$ is based on SES and therefore depends on the level $\ell_t$ and the innovations, we will use the notation $R_{t+h|t}$ hereafter for it and any quantity derived from it. As pointed out above, we my assume by the central limit theorem that the aggregate of the remaining-course dose metric $R_{t+h|t}$, is approximately normally distributed with mean $\widehat{R}_{t+h|t} = h\ell_t$ and variance $\sigma^2\Omega(h)$, then the central $(1-\gamma)\cdot 100\%$ prediction interval (PI) is given by (cf. Hyndman et al. [14], section 6.5 p. 88)):

$$PI = h\ell_t \quad \pm \; z_{\gamma/2}\,\hat{\sigma}_t\sqrt{\Omega(h,\alpha)}, \tag{13}$$

where $\hat{\sigma}_t^2$ is the exact maximum likelihood estimate of the innovation variance $\sigma^2$, which can be computed from the residuals (innovations) up to fraction $t$ as follows,

$$\hat{\sigma}_t^2 = \frac{1}{t - \mathrm{K} - \mathrm{M}} \sum_{s=1}^{t} \varepsilon_s^2 \tag{14}$$

where K denotes the number of parameters estimated in the state-space model and M is the number of missing residuals. In the present study, $\alpha$ was prespecified at each model evaluation rather than estimated; only $\hat{\ell}_0(t,\alpha)$ was estimated. Therefore, $\mathrm{K} = 1$ for the local-level model considered here (see Equation (3)). For more complicated homoscedastic linear state-space models, K should likewise equal the number of parameters actually estimated. In the retrospective adaptive-monitoring analysis, all residuals were known, so $\mathrm{M} = 0$.

In other applications such as prospective online adaptive monitoring, the most recent residual for fraction $t$ will not be available since it is yet to be delivered, so one may choose to simply set $\varepsilon_t \equiv 0$ in equation (14) summing up to $t$, but incrementing M by one to $\mathrm{M} = 1$. This points the way of how to deal with $j$ missing residuals, simply set them to zero in the above sum and set $\mathrm{M} = j$. Observe that dividing by the degrees of freedom $t - \mathrm{K} - \mathrm{M}$ rather than the sample size $t$ in equation (14) makes this estimate less biased (cf. the discussion following equation (36) in Ref. [17]).

### 2.5 Analytic Gaussian Predictive Distribution

Denote by:

$$S_t = \sum_{i=1}^{t} y_i \tag{15}$$

the aggregate of the observed dose metric through fraction $t$. Then the aggregate of the estimated final course scalar dose metric is given by:

$$M_{N|t} = S_t + R_{t+h|t}, \tag{16}$$

and the aggregate of the final estimated fraction-averaged metric is

$$\bar{y}_N = \frac{M_{N|t}}{N}. \tag{17}$$

The same predictive distribution can therefore be expressed either for the aggregate of the remaining estimated scalar dose metric, the aggregate of the final estimated scalar dose metric, or the aggregate of the final estimated average scalar dose metric through a location shift and, for the average, a scale factor of $1/N$.

To quantify the likelihood that the aggregate of the estimated final scalar dose metric stays below a clinically relevant threshold, we use the same normal approximation that underpins the prediction intervals above. Under SES with approximately independent, homoscedastic, Gaussian innovations, the aggregate of the remaining-course scalar dose metric satisfies

$$R_{t+h|t} \approx \mathcal{N}\left(h\ell_t, \hat{\sigma}_t^2 \Omega(h, \alpha)\right). \tag{18}$$

Since, $R_{t+h|t}$ is assumed to be approximately normally distributed with mean $h\ell_t$ and variance $\hat{\sigma}_t^2\Omega(h)$, for any dose metric threshold $c_{\text{thr}}$ on $R_{t+h|t}$ one has

$$P\Big(R_{t+h|t} \leq c_{\text{thr}}\Big) \approx \Phi\left(\frac{c_{\text{thr}} - h\ell_t}{\hat{\sigma}\sqrt{\Omega(h, \alpha)}}\right), \tag{19}$$

where $\Phi$ denotes the standard normal cumulative distribution function. For a prespecified threshold $C_{\text{thr}} = S_t + c_{\text{thr}}$ on the aggregate of the final dose metric, one then finds using equations (5) and(16) the following predictive probability:

$$P\Big(M_{N|t} \leq C_{thr}\Big) \approx \Phi\left[\frac{C_{thr} - (S_t + h\ell_t)}{\hat{\sigma}_t\sqrt{\Omega(h, \alpha)}}\right], \tag{20}$$

This quantity is a model-based conditional probability, not a validated clinical likelihood nor validated decision rule. The clinical meaning of $C_{thr}$ and any probability threshold for action must be established separately in the decision layer.

The analytic interval treats $\alpha$ and $\hat{\sigma}_t^2$ as fixed plug-in quantities and does not include parameter-estimation uncertainty. It also represents temporal variation in the observed dose metric time series; it does not separately propagate contouring, registration, synthetic-CT, or dose-calculation uncertainty.

## 2.6 Retrospective Data and Dose Metric Construction

Time-series data were obtained retrospectively from fraction-dose calculations for 32 prostate and 19 head-and-neck radiotherapy courses. MIM ART Assist (MIM Software Inc., Cleveland, Ohio, United States) was used to manage daily cone-beam computed tomography (CBCT) images and subsequent image processing. Synthetic CT images were generated from CBCT, planning-CT contours were propagated, and fractional dose was recalculated on the synthetic CT using MIM SureCalc with the SciMoCa Monte Carlo dose engine commissioned to the clinical linear accelerator beam models. DVH data were exported from each fractional dose distribution and processed with an in-house Python pipeline.

A per-fraction mean-dose time series was extracted for each OAR listed in Table 1. Not all structures were contoured for every patient; the resulting dataset comprised 340 OAR time series. Multiple OAR time series were nested within each treatment course.

Table 1: Organ-at-risk (OAR) time series used for the retrospective interval characterization.

| **Treatment site** | **OAR time series included** |
|---|---|
| Prostate | Rectum (32), bladder (32), sigmoid (31), & left and right femoral head (32/32) |
| Head and neck | Left and right parotid (22/22), spinal cord (22), larynx (23), mandible (23), left and right submandibular gland (15/15), esophagus (21), & oral cavity (18) |

### Empirical Interval Characterization

The purpose of this analysis was to characterize interval behavior across forecast horizons and LT-SES temporal-response settings relative to a running-mean benchmark. Each OAR time series was evaluated at candidate $\alpha$ values from 0.1 to 1.0 in increments of 0.1, and the forecast horizon $h$ ranged from 5 to 20 remaining fractions in five-fraction increments. Results for each candidate $\alpha$ setting were then summarized across time series. A running-mean benchmark was evaluated separately using the mean of the observations available through the forecast origin as the non-updating level, which as discussed above is akin to considering the boundary case of $\alpha = 0$. For each complete time series of length $N$ and each eligible horizon, the forecast origin was

$$t = N - h. \tag{21}$$

The data available through $t$ were used to estimate the remaining-course interval for each configuration. Because the OAR time series had different physical scales, interval width was normalized for cross-series comparison. At each forecast origin, the projected final aggregate was

$$\widehat{M}_{N|t}(\alpha, h) = S_t + h\ell_t(\alpha). \tag{22}$$

The time series, the realized remaining aggregate, and interval endpoints were divided by the positive quantity $\widehat{M}_{N|t}$. The normalized projected final aggregate therefore equaled 1 by construction, and normalized interval width was $(U - L)/\widehat{M}_{N|t}$. Positive rescaling does not change interval coverage; normalization was used only to compare widths across physical scales. This scaling does not create a deformably accumulated dose distribution.

For each OAR time series, configuration, and forecast horizon, it was recorded wether the realized remaining aggregate was contained in the nominal 95% or 99% prediction interval. Empirical coverage was the proportion of included time series for which the interval contained the realized value. Coverage was evaluated for the complete dataset and separately for the head-and-neck and parotid cohorts. Because several OAR time series arose from the same treatment course, these percentages are descriptive time series-level summaries rather than independent patient-level inference.

### Illustrative Parotid Monitoring Examples

Two right-parotid mean-dose time series from head-and-neck courses were selected to contrast differing patterns of interfractional variation. Both exhibited a substantial level shift around fraction 20, with greater day-to-day variation in the first time series than in the second.

If $y_t$ is the physical fraction level mean dose delivered up to fraction $t$ fractions and $N$ is the number of fractions, the plotted course-equivalent daily value was $z_t = Ny_t$. The final fraction-averaged course-equivalent metric is

$$\bar{z}_N = \frac{1}{N}\sum_{t=1}^{N} z_t = \sum_{t=1}^{N} y_t. \tag{23}$$

This scaling places the fraction-specific values and projected final metric on a consistent course-dose axis. It does not create a deformably accumulated dose distribution.

At each displayed fraction, the analytic Gaussian model was used to calculate the projected final metric, its 95% prediction interval, and the conditional probability of finishing below each illustrative threshold. The thresholds were set 1% below and 1% above the final observed course-equivalent metric to create contrasting eventual fail/pass behavior. LT-SES was evaluated at $\alpha = 0.25, 0.50$, and 1.0 to illustrate a range of temporal-response settings, from greater stability against short-term variation to greater responsiveness to recent observations. Probability curves were displayed beginning at the first forecast origin for which five one-step innovations were available after initialization.

## 2.7 Illustrative Monitoring Outputs

### Threshold Color Map

Traffic-light protocols have been used in radiotherapy to identify patients requiring adaptive review or treatment modification, and similar color coding has been applied to deviations from OAR dose-volume constraints [18, 19]. Adapting this visual format to a probabilistic output, the conditional probability of meeting the final-course threshold was grouped, for illustration, into four nonoverlapping levels: replace with: $P < 0.01$, $0.01 \leq P < 0.50$, $0.50 \leq P \leq 0.99$, and $P > 0.99$, displayed using a sequential dark-blue-to-yellow color scale. The maps evaluate a range of thresholds around the final observed mean so that model response can be examined across fractions and smoothing parameters. A positive threshold difference denotes a threshold above the final observed mean; a negative difference denotes a more stringent threshold below it. This range is shown only to characterize model behavior. In clinical use, a prespecified threshold would be applied to each monitored dose metric, yielding a single monitoring classification after each fraction.

### Rank Scoring of Scalar Dose Metrics

Quantitative scoring methods have been used in medical physics and radiation oncology to rank treatment plans, and more recently to identify changes in plan quality associated with adaptive replanning [20, 21]. Extending this general approach to fraction-level monitoring, the following rank score at fraction $t$, $RS(t)$, was calculated after each fraction to prioritize individual scalar dose metrics for review:

$$RS(t) = \left[1 - P(\widehat{M}_{N|t} \leq C_{thr})\right] \frac{\widehat{M}_{rem}(t)}{M_{req,rem}(t)}, \tag{24}$$

where $P(\widehat{M}_{N|t} \leq C_{thr})$ is the probability that the final-course mean dose metric $\widehat{M}_{N|t}$ will be less or equal to the selected threshold (cf. equations (20) and (22)), $\widehat{M}_{rem}(t) = \widehat{R}_{t+h|t}/h = \ell_t$ is the forecast mean dose metric over the remaining fractions, and $M_{req,rem}(t)$ is the remaining mean dose metric required to finish at the threshold. The required remaining mean dose metric was calculated as

$$M_{req,rem}(t) = \frac{C_{thr} - \sum_{i=1}^{t} y_i}{N - t}, \tag{25}$$

where $N$ is the total number of fractions, $C_{thr}$ is the selected final-course dose metric threshold, and $y_i$ is the observed dose metric at fraction $i$. Note that $1 - P(\widehat{M}_{N|t} \leq C_{thr})$ in equation (24) is the probability that the final mean dose metric will exceed the threshold $C_{thr}$. Higher scores therefore indicate both a greater likelihood of exceeding the threshold and less opportunity for the course to recover. The score is intended as a relative review-prioritization index rather than a calibrated clinical-risk measure.

To isolate the contribution of probability weighting from a simple scalar comparator, the rank score was evaluated alongside the running-mean-to-required-mean ratio

$$Q(t) = \frac{\overline{y}_{1:t}}{M_{req,rem}(t)}, \qquad \overline{y}_{1:t} = \frac{1}{t}\sum_{i=1}^{t} y_i. \tag{26}$$

A value of $Q(t) = 1$ indicates that the running mean equals the mean required over the remaining fractions to finish at the threshold; values below or above unity indicate that the running mean is below or above that requirement, respectively.

Scores were evaluated from fraction 5 onward using $\alpha = 1.00$, 0.50, and 0.25. For each parotid time series, the illustrative threshold was fixed at 5% above the running mean available at fraction 5, $T = 1.05\,\overline{y}_{1:5}$, and was then held constant for all subsequent fractions. This early-course reference allowed the scalar ratio and probability-weighted rank score to be compared without defining the rank-score threshold from the eventual final observed mean.

A special case occurs when $M_{req,rem}(t) \leq 0$. At that point, the threshold cannot be met even if all remaining fraction-level values are zero, so the ratio is not meaningful. Such metrics should be flagged separately and excluded from numerical ranking. In addition, the ratio can become unstable as a positive denominator approaches zero; any operational implementation would require a prespecified minimum denominator, cap, or separate near-zero flag.

# 3 Results

## 3.1 Prediction-Interval Width

As shown in Figure 2, mean normalized 95% prediction-interval width increased with forecast horizon for all evaluated configurations and increased monotonically with $\alpha$ among the LT-SES configurations. At $h = 5$, mean widths ranged from approximately 0.01 for the running-mean benchmark to approximately 0.03 for $\alpha = 1.0$. At $h = 20$, the corresponding range was approximately $0.02 - 0.23$. Among the LT-SES configurations, larger values of $\alpha$ increased the propagated remaining-course variance through the lead-time factor $\Omega(h, \alpha)$ (see Equation (9)), producing wider model-based prediction intervals, particularly at longer horizons.

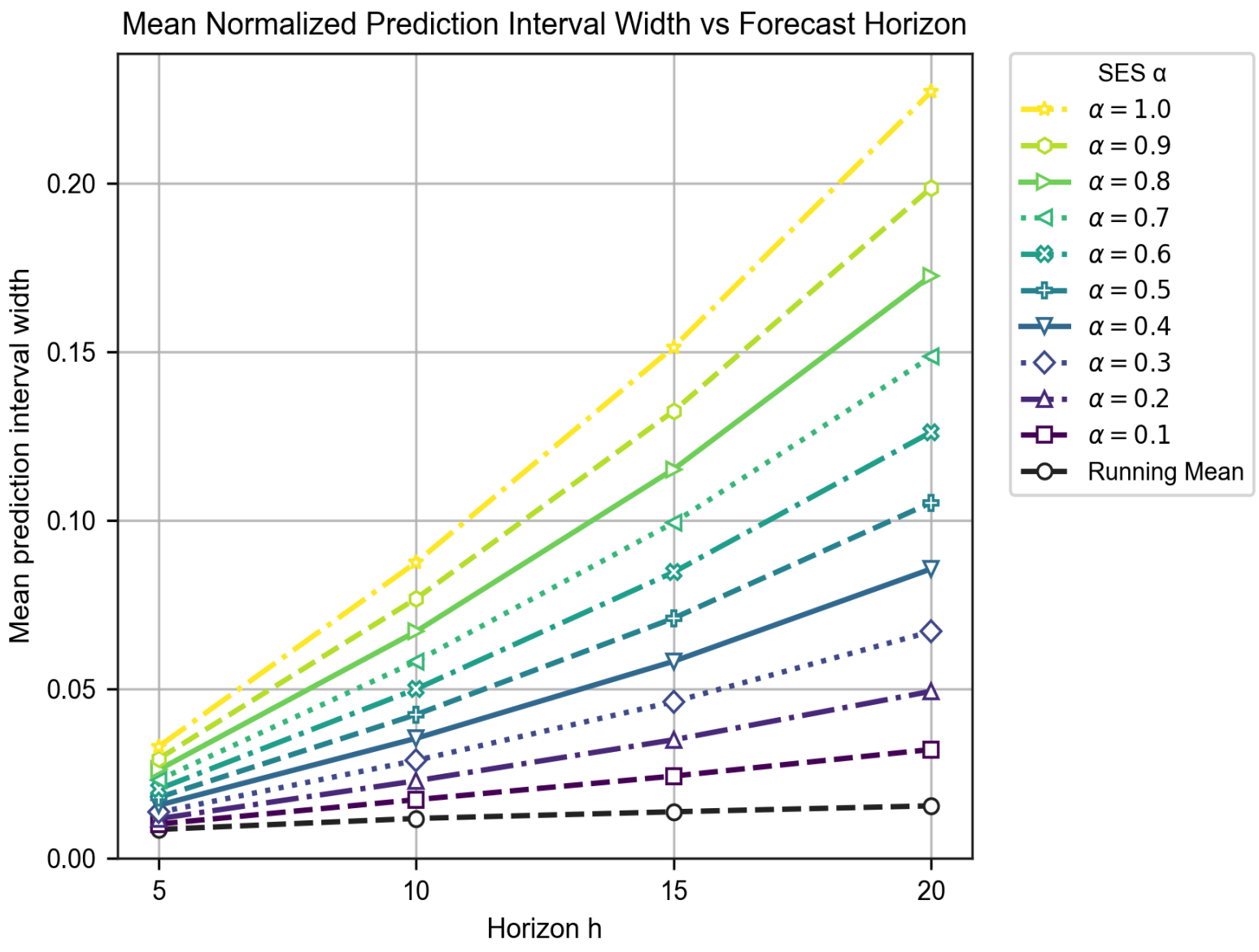

Figure 2: Mean normalized 95% prediction-interval width as a function of remaining-course horizon for a running-mean benchmark and LT-SES $\alpha$ settings from 0.1 to 1.0. Width increased with horizon for all settings and with $\alpha$ among the LT-SES settings.

## 3.2 Empirical Coverage

Empirical coverage differences were concentrated in the running-mean benchmark and the lower $\alpha$-values (see Figure 3). For the nominal 95% intervals, coverage for the running-mean benchmark decreased from approximately 0.75 at $h = 5$ to 0.39 at $h = 20$, while coverage at $\alpha = 0.1$ it decreased from approximately 0.88 to 0.74. Undercoverage was smaller for $\alpha = 0.2$, while for $\alpha = 0.3$ it remained near the nominal level across the evaluated horizons. Values of $\alpha \geq 0.4$ produced coverage of approximately $0.98 - 1.0$ across the evaluated horizons. The nominal 99% intervals showed the same general pattern, with greater variation in the smaller parotid cohort.

For the running-mean benchmark, the forecast level is the mean of the observations available through the forecast origin, and $\hat{\sigma}_t$ is the standard deviation of observations about that mean. Its undercoverage relative to the LT-SES configurations is consistent with the benefit of allowing the local level to evolve. However, increasing $\alpha$ also mechanically widens the intervals through $\Omega(h, \alpha)$ (see Equation (9)), so coverage alone cannot separate improved tracking from increased interval width. Therefore, coverage analysis characterizes prediction-interval behavior in all evaluated temporal-response configurations and does not establish a uniquely appropriate value of $\alpha$.

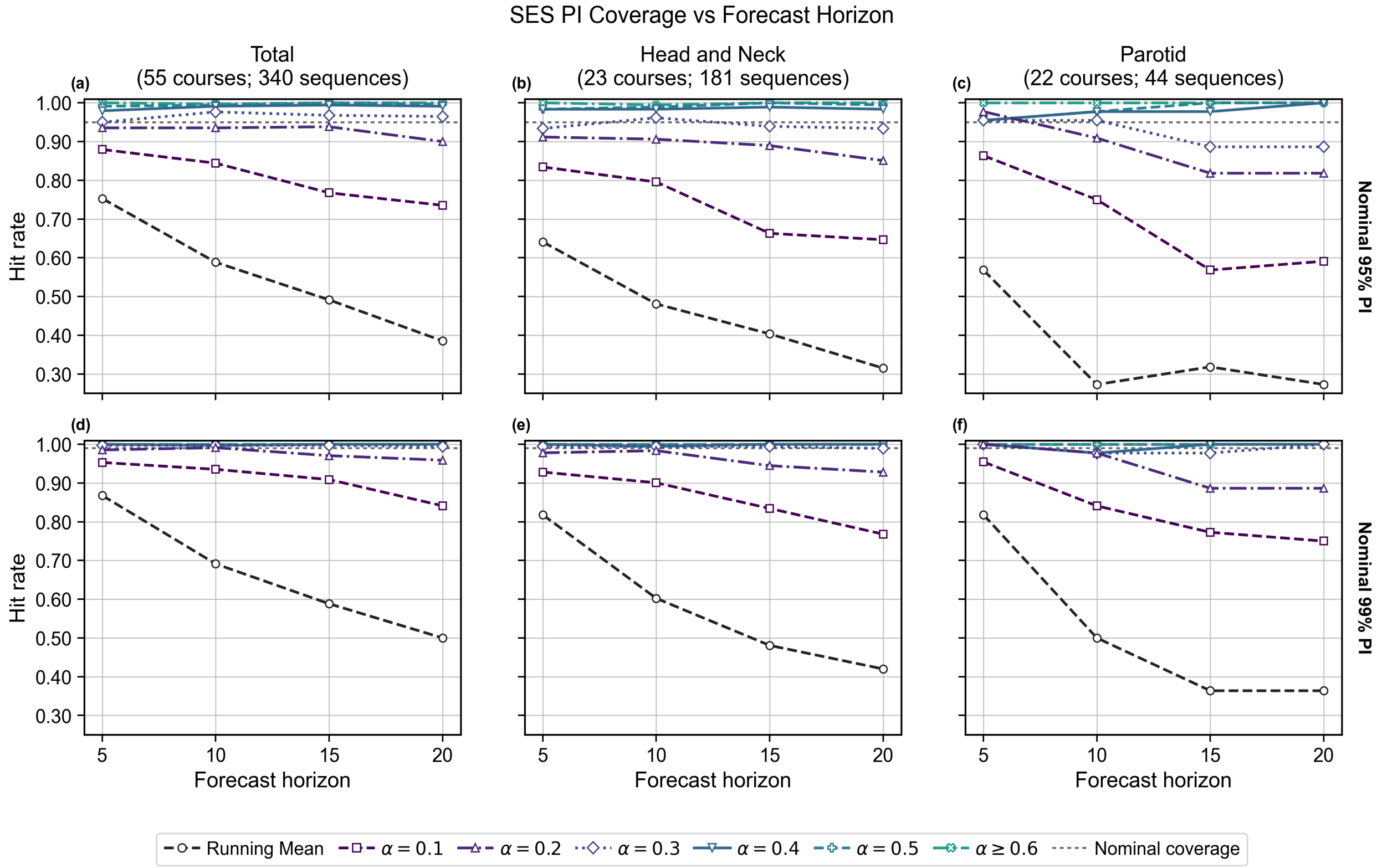

Figure 3: Empirical coverage of the nominal 95% and 99% analytic Gaussian prediction intervals for the complete dataset, head-and-neck cohort, and parotid cohort across forecast horizons for a running-mean benchmark and LT-SES $\alpha$ settings from 0.1 to 1.0. The curves for $\alpha = 0.6 - 1.0$ overlapped at the plotted scale and are represented collectively in the legend as $\alpha \geq 0.6$. Dashed lines indicate nominal coverage. The plotted proportions are descriptive time series-level summaries.

## 3.3 Illustrative Parotid Examples

The two parotid time series demonstrated the effect of $\alpha$ under contrasting patterns of interfractional variation (see Figures 4 and 5). HN3 showed substantial interfractional variation, while HN15 exhibited less short-term fluctuation and a clearer sustained increase beginning near the middle of the treatment course. In both time series, $\alpha = 0.25$ produced a smoother, more gradual response, $\alpha = 1.0$ responded most strongly to recent observations, and $\alpha = 0.50$ provided an intermediate response.

These differences propagated to the projected final-course means and threshold probabilities. In HN3, higher $\alpha$ produced greater interfractional responsiveness, whereas lower $\alpha$ suppressed short-term variation. In HN15, higher $\alpha$ responded more rapidly to the sustained level shift, while lower $\alpha$ adapted more gradually. Differences among configurations were greatest while substantial treatment remained and generally diminished as additional observations accumulated. Together, the examples illustrate the configurable tradeoff between stability against short-term variation and responsiveness to persistent change.

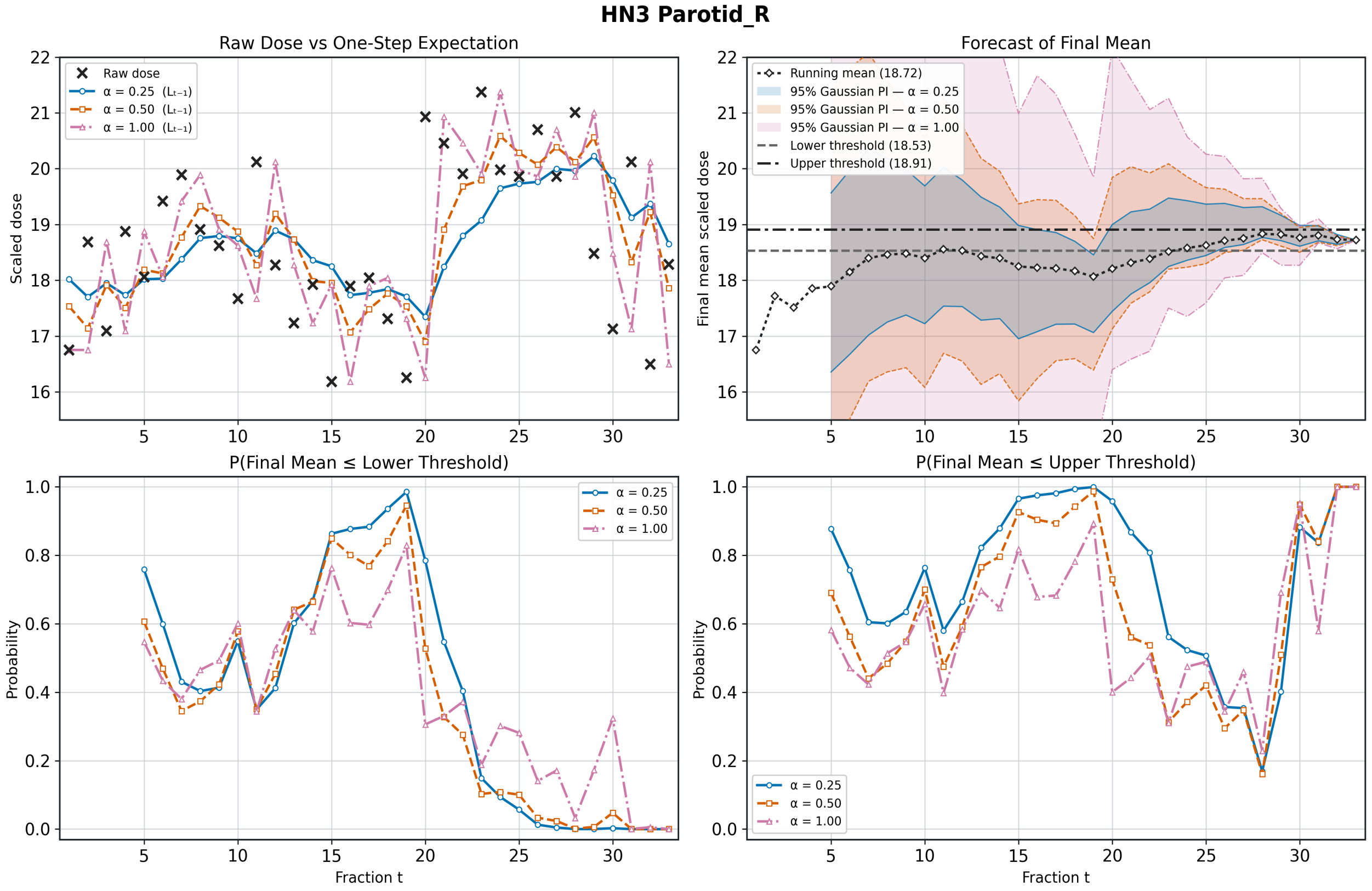

Figure 4: Illustrative HN3 right-parotid time series with substantial interfractional variation. The upper-left panel shows the course-equivalent daily metric and LT-SES level estimates. The upper-right panel shows the running mean together with the $\alpha$-specific 95% prediction intervals for the projected final mean. Lower panels show the corresponding probabilities of meeting the illustrative lower and upper thresholds.

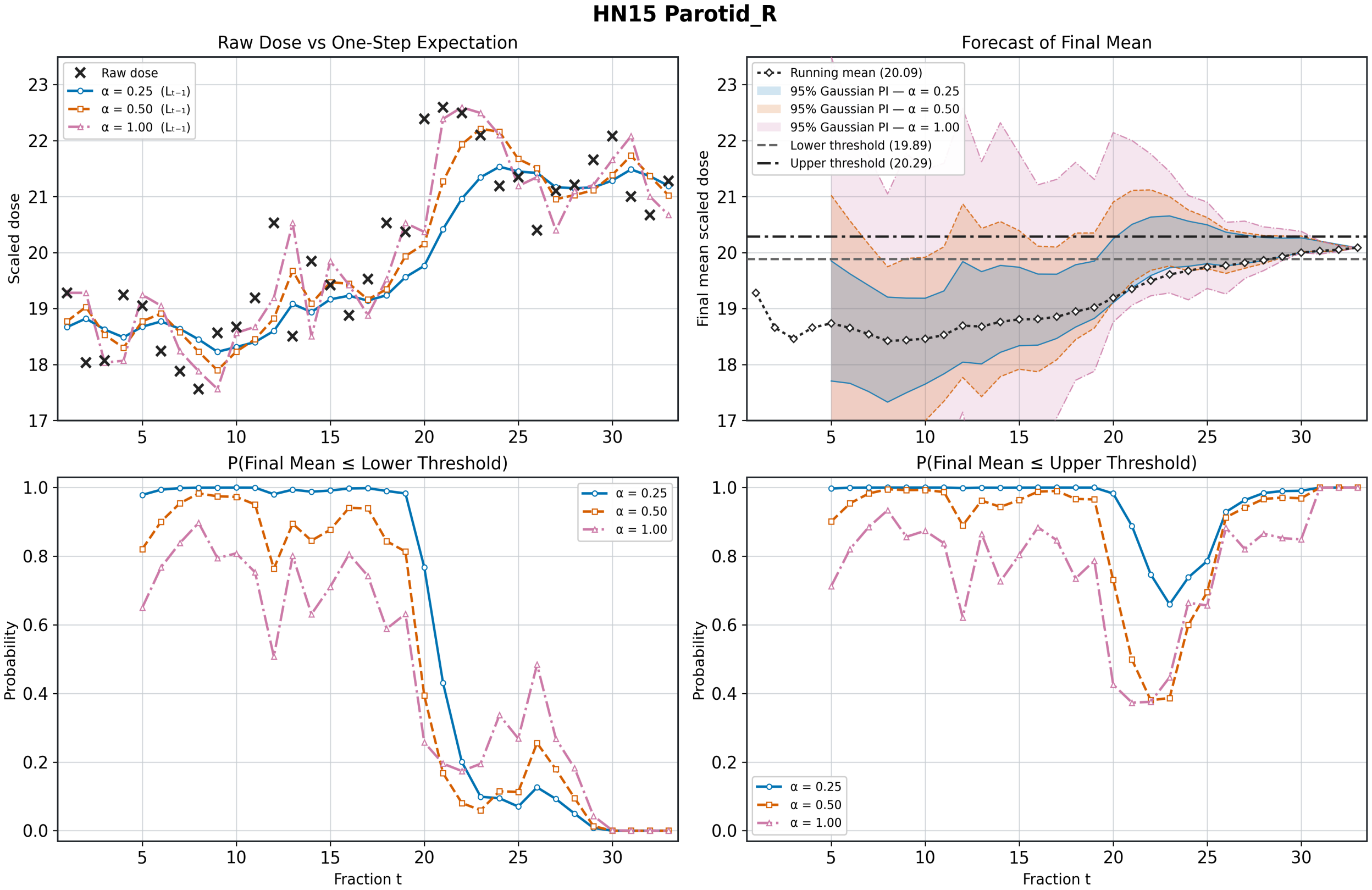

Figure 5: Illustrative HN15 right-parotid time series with lower short-term variability and a sustained midcourse increase. Panels are defined as in Figure 4. Higher values of $\alpha$ respond more rapidly to the level change, whereas lower values adapt more gradually.

### 3.4 Initial-Level Estimation

Figure 6 shows the estimated initial level $\widehat{\ell}_0(t, \alpha)$ as progressively more observations were included. For $\alpha = 0.50$ and 0.25, $\widehat{\ell}_0$ was re-estimated at each forecast origin and stabilized as additional fractions accumulated. The greatest early dependence on forecast origin occurred for $\alpha = 0.25$. In HN3, $\widehat{\ell}_0$ increased during the early fractions before stabilizing near 18.0, whereas in HN15 it initially decreased and then gradually approached approximately 18.7.

The special case $\alpha = 1.0$ differs. By definition, $\widehat{\ell}_0 \equiv y_1$ (see Section 2.3), so the estimate is determined by the first observation and remains unchanged at every forecast origin.

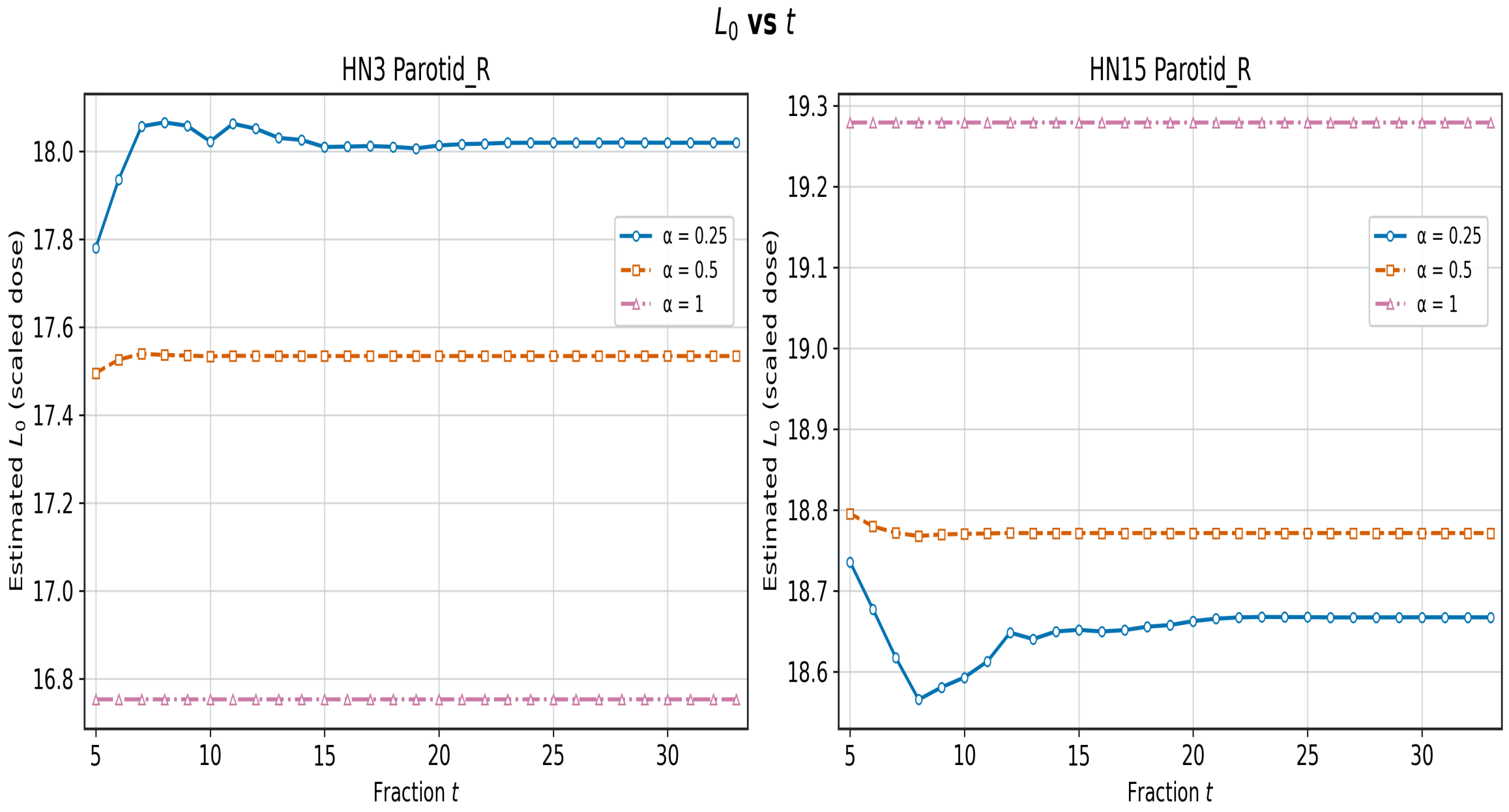

Figure 6: Estimated initial level $\widehat{\ell}_0(t, \alpha)$ as a function of forecast origin $t$ for HN3 and HN15 right-parotid time series using $\alpha = 1.0$, 0.50, and 0.25. For $\alpha = 1.0$, $\widehat{\ell}_0 = y_1$ and is constant; for $\alpha = 0.50$ and 0.25, the estimate changes as additional observations become available before stabilizing.

### 3.5 Threshold Color Map

The probability maps illustrate the tradeoff between responsiveness and stability when selecting $\alpha$ (see Figure 7). With $\alpha = 1.0$, the classification responds quickly to recent changes, but varies more between fractions. Smaller values of $\alpha$ produce smoother, more persistent regions, although sustained changes appear more gradually.

The vertical position of a prespecified threshold relative to the final mean can be determined only retrospectively. During treatment, the final mean is unknown; therefore, the operational output would be the evolving category for a single predefined threshold. The full map is presented to show how that category would have differed had the threshold been more or less stringent. Consequently, the colored regions displayed show how the model output changes between thresholds, fractions, and smoothing parameters.

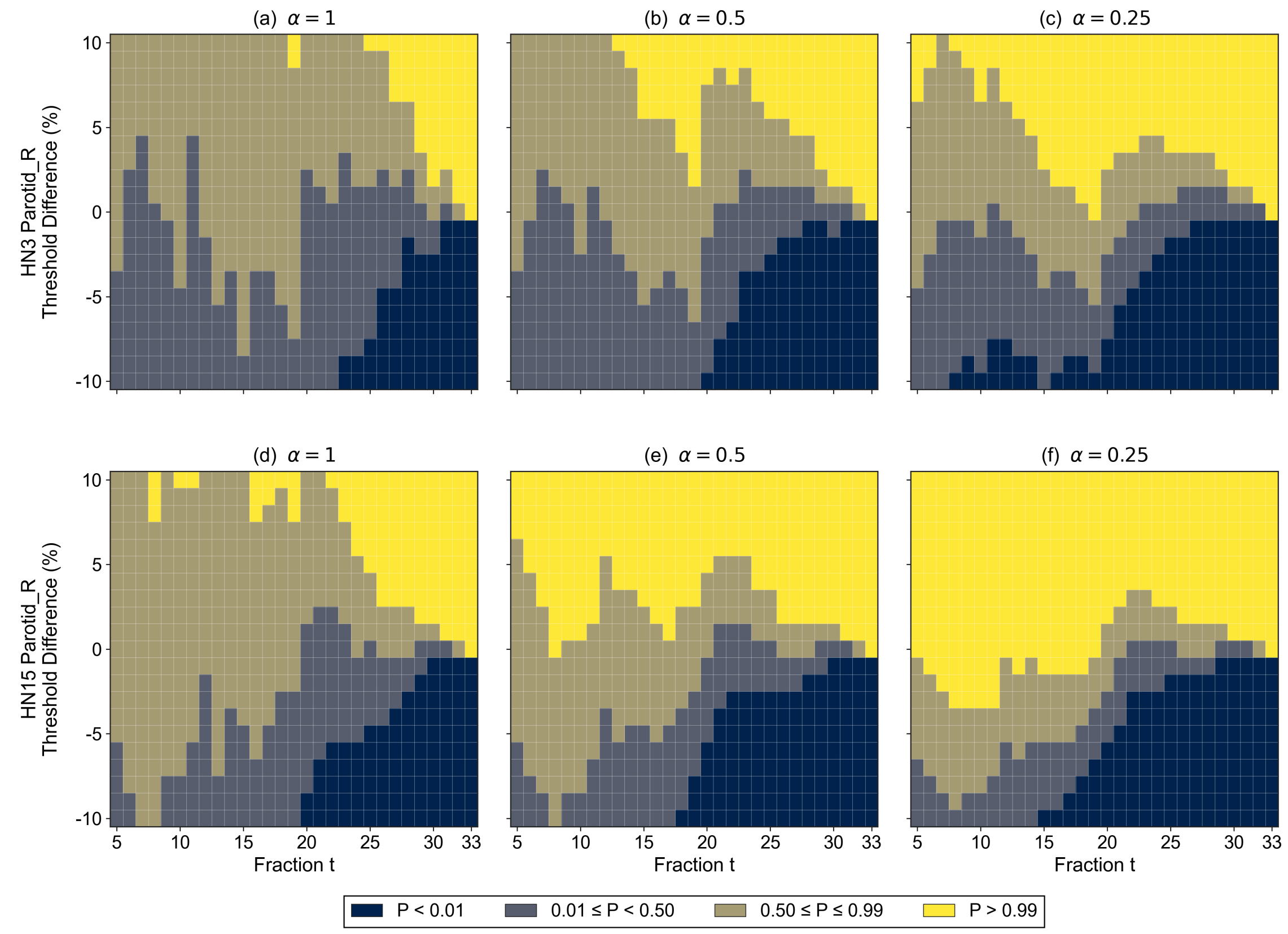

Figure 7: Four-level probability maps for Parotid_R in HN3 (a–c) and HN15 (d–f) using $\alpha = 1.0$, 0.5, and 0.25. Colors progress from dark blue to yellow with increasing probability that the final course-averaged metric will satisfy the evaluated threshold: $P < 0.01$, $0.01 \leq P < 0.50$, $0.50 \leq P \leq 0.99$, $P > 0.99$. Threshold difference is expressed relative to the final observed mean.

### 3.6 Rank Scoring of Dose Metrics

For the rank-score comparison, each time series-specific threshold was fixed at 5% above the running mean at fraction 5 (see Figure 8). Through treatment fraction 20 the unweighted running-mean-to-required-mean ratios $Q(t)$ remained similar and dropped below unity to approximately 0.925 for HN3 and 0.940 for HN15, a difference of only 0.015. Incorporating the probability of threshold exceedance produced substantially greater separation at treatment fraction fraction 20. The rank scores $RS(t)$ for HN3 and HN15 for $\alpha = 1.0$ were approximately 0.62 and 0.86, for $\alpha = 0.50$ were close to 0.32 and 0.78 , and for $\alpha = 0.25$ were about 0.08 and 0.50, respectively. Decreasing $\alpha$ reduced short-term variation and delayed the response, but HN15 received the higher score at fraction 20 for all three settings and remained higher thereafter. Note that the HN15 scalar ratio did not reach unity until approximately fraction 24, indicating that the probability-weighted score distinguished the sustained trajectory change several fractions before the scalar comparator crossed unity while preserving the principal post-transition ordering across the tested smoothing parameters.

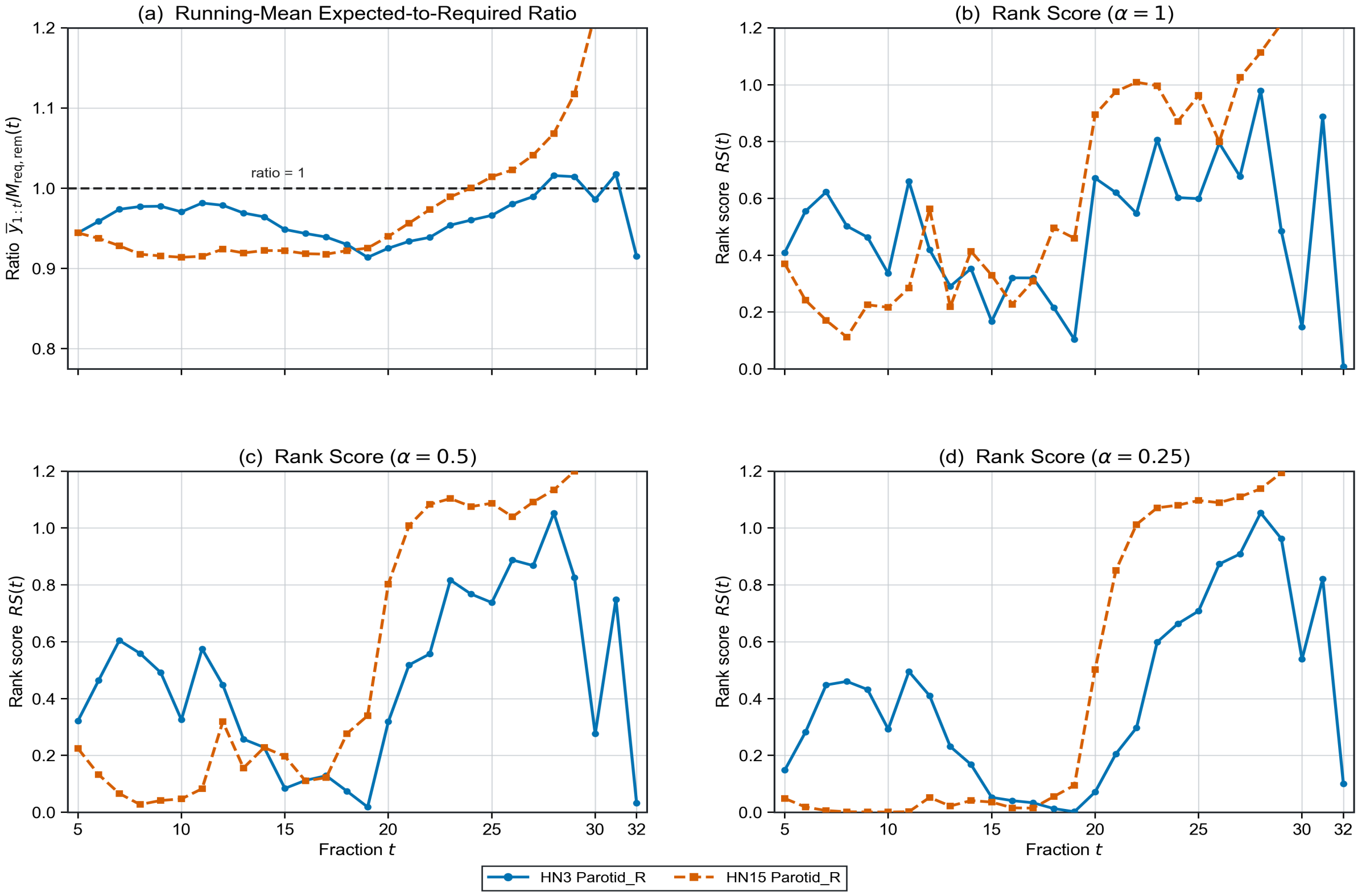

Figure 8: Comparison of a conventional scalar running-mean-to-required-remaining-mean ratio with probability-weighted rank scores for the HN3 (blue) and HN15 (orange) right parotid time series. Panel (a) shows $Q(t) = \bar{y}_{1:t}/M_{req,rem}(t)$; the dashed line at $Q(t) = 1$ denotes equality between the observed running mean and the remaining mean required to finish at the threshold. Panels (b)–(d) show $RS(t)$ for $\alpha = 1.0$, 0.50, and 0.25, respectively. For each time series, the threshold was fixed at 5% above its running mean at fraction 5 and held constant thereafter. The rank score weighs the LT-SES expected-to-required remaining-mean ratio by the model-based probability of threshold exceedance, $1 - P(t)$. Values above 1.2 are clipped by the plotting limit.

## 4 Discussion

This work introduces remaining-course dose-metric analysis as a lead-time forecasting problem. Given a sequential scalar metric generated by an ART measurement pipeline, LT-SES provides a current local-level estimate and a transparent mechanism for propagating future innovations into the aggregate of the unobserved fractions, supplementing a projected final metric with a predictive distribution for the unobserved fractions. The projected threshold difference in Equation (20) is thereby interpreted in the context of time series-specific interfractional variation and remaining-course horizon. The smoothing parameter $\alpha$ provides a configurable temporal response for a monitored OAR/PTV time series: smaller values place greater persistence on the current level of the time series and attenuate short-term fluctuations, whereas larger values give greater weight to recent observations and respond more rapidly to persistent change. Selection of $\alpha$ should therefore be tied to the clinical monitoring objective, including the need for rapid response versus the degree of confidence that the current level will persist. The current level, innovation variance, threshold, and remaining-course requirement remain time series specific.

The retrospective interval analysis characterized this configurability rather than identifying an optimal $\alpha$. Empirical-coverage differences were concentrated in the running-mean benchmark and lower $\alpha$-values and at longer horizons, while interval width increased with $\alpha$ through the lead-time variance among the LT-SES settings. Because empirical coverage and interval width change together, these results describe the behavior of the evaluated settings rather than establish a preferred $\alpha$-value. Selection of $\alpha$ should therefore reflect the intended monitoring response and should not be based on empirical coverage alone. The analysis applied the same candidate grid to each time series for descriptive comparison, but it did not test whether a setting selected for one metric or treatment context transfers to another. Metric- or application-specific selection should therefore be prespecified and clinically justified.

The two parotid time series provide proof-of-concept evidence of configurable behavior. Their contrasting interfractional patterns demonstrate the expected stability–responsiveness tradeoff in the level estimates, threshold probabilities, probability maps, and rank scores. In the rank-score example, the probability-weighted index produced earlier and greater numerical separation than the selected scalar comparator while preserving the principal post-shift ordering. These examples show that configurability remains evident in threshold-relative monitoring and review-prioritization outputs.

Standardization by the estimated predictive scale expresses each metric relative to its own threshold and observed variation. The presented examples illustrate relative review prioritization, but they do not establish that a specific $\alpha$-value transfers between metrics. The resulting probabilities are model-based monitoring indices whose numerical calibration depends on the assumed innovation model. Departures from independent, constant-variance Gaussian innovations, including autocorrelation, changing variance, nonlinear drift, abrupt regime changes, or parameter-estimation uncertainty, may affect nominal interval coverage and probability values. Metric-specific thresholds, clinical priorities, probability categories, ranking criteria, and associated actions therefore remain part of the decision layer.

The retrospective coverage analysis was descriptive: multiple OAR time series were nested within treatment courses and one forecast origin was evaluated per time series and horizon. These features limit population-level generalization but do not alter its intended use here to characterize LT-SES configuration behavior. Measurement-layer uncertainties from imaging, contouring, registration, synthetic CT, and dose calculation are not separately modeled by LT-SES. Random measurement variation may contribute to the estimated innovation scale, whereas systematic or correlated errors are not explicitly represented. An additional margin may be applied to the monitoring threshold when clinically appropriate, although such a margin does not explicitly model these uncertainty sources.

When few innovations are available, $\hat{\sigma}_t^2$ may be unstable, so probability and standardized-margin outputs should be displayed only after a prespecified minimum residual count. In this study, probability curves were shown once five residuals were available following initialization. In our practice, 15 fractions is used as the lower course-length limit for LT-SES monitoring. LT-SES may not be suitable for ultrahypofractionated treatments, for which the small number of fractions leaves limited data for innovation-scale estimation and sequential updating. The framework could in principle be adapted to conventionally fractionated treatments delivered in sequential phases or with cone-down plans by maintaining phase-specific metric time series and combining their projected contributions. Such an extension would require explicit treatment of cross-phase covariance when plan-specific innovations are correlated.

The present evaluation primarily concerns fraction-averaged daily-anatomy metrics. These provide an additional safety-monitoring layer beyond reliance on the initial plan but do not preserve the spatial history of dose deposition or represent reference-volume accumulated dose. The LT-SES framework is not intrinsically limited to tier-2 metrics and could also be applied to tier-3 accumulated-dose metrics, including voxel-level time series.

Determining an optimal $\alpha$-value for a particular monitoring objective is outside the scope of the present work. Within the proposed framework, however, $\alpha$ is a decision-layer configuration parameter that may be selected prospectively for the monitored dose metric or application according to the desired temporal responsiveness and the degree of confidence that the current level will persist. Supplementary Section S3 provides additional model comparisons and alternative approaches, including cohort- and series-specific fitting, higher-order exponential-smoothing models, and an exchangeable Bayesian reference, which may inform decision-layer selection of the monitoring configuration.

# 5 Conclusion

LT-SES provides a practical and transparent framework for projecting a final treatment-course scalar dose metric while representing temporal uncertainty from fractions that remain unobserved. For a given monitored metric and clinical objective, $\alpha$ provides a prospectively specified temporal-response setting: higher values favor rapid response to recent observations, whereas lower values favor greater persistence of the current level estimate. The projected scalar difference from a threshold is expressed relative to the time series-specific predictive variation and remaining-course horizon, providing an additional uncertainty-aware layer of information for review prioritization. In the two illustrative parotid time series, the expected $\alpha$-dependent behavior remained evident in threshold probabilities, probability maps, and probability-weighted rank scores, providing proof-of-concept evidence that the configurable temporal response propagates to interpretable monitoring and review-prioritization outputs. The resulting probability-scale outputs and rank scores complement conventional scalar monitoring under prespecified, clinically justified monitoring configurations.

# Supplementary Material

## Lead-Time Simple Exponential Smoothing as a Configurable Framework for Adaptive Radiotherapy Monitoring

Robert Boyd and Wolfgang A. Tomé

# Introduction

The following material provides mathematical details and exploratory analyses that extend the main manuscript while preserving its focus on lead-time simple exponential smoothing (LT-SES) as a configurable framework for adaptive radiotherapy monitoring. Section S1 develops the SES–ARIMA representation and the propagation of future innovations into the remaining-course aggregate. Section S2 describes the empirical residual-bootstrap construction and its fast Fourier transform (FFT) implementation. Section S3 presents a descriptive predictive comparison of SES with alternative forecasting models. Section S4 summarizes our institutional experience and practical considerations that motivated the LT-SES workflow, including dose-metric selection, selection of the smoothing parameter, and the use of probability-based prioritization for adaptive review. Displayed equations are numbered independently by section as (S1.x), (S2.x), and (S3.x); tables and figures use the corresponding section-based numbering.

# Table of Contents

# Supplementary Section S1: SES, ARIMA, and Lead-Time Innovation Propagation

## S1.1. ARIMA Representation of the Fraction-Level Metric

The observed dose metric process $\{y_t\}$ has been modeled using SES with level $\ell_t$ and smoothing parameter $\alpha$, and forecasts and prediction intervals have been constructed for the aggregate of the dose metric by summing the point forecasts. It is now shown that this forecasting scheme is the optimal predictor for a nonstationary ARIMA(0,1,1) dose metric process $\{y_t\}$ whose level is allowed to fluctuate and an ARIMA(0,2,1) stochastic process for the nonstationary aggregate $\{R_t\}$ of the dose metric process $\{y_t\}$ .

Let $\{z_t\}$ be an arbitrary nonstationary stochastic process, $B$ the backshift operator defined by the functional equation $Bz_t = z_{t-1}$, and $\{\varepsilon_t\}$ independent innovations that are identically distributed with mean 0 and variance $\sigma^2$. The linear stochastic process $\{\Delta_t\}$ given by

$$\Delta_t = \mu + \varepsilon_t + \psi_1\varepsilon_{t-1} + \psi_2\varepsilon_{t-2} + \cdots$$

is stationary if $\sum_{i=0}^{\infty} |\psi_i| < \infty$. A nonstationary stochastic process $\{z_t\}$ for which the *d-th* difference of the process is stationary can be represented by the following model:

$$\phi(B)\Delta_t = \theta(B)\varepsilon_t, \text{ where } \Delta_t \equiv (1-B)^d z_t \tag{S1.1}$$

And

$$\phi(B) \equiv 1 - \phi_1 B - \phi_2 B^2 - \cdots - \phi_p B^p \tag{S1.2}$$
$$\theta(B) \equiv 1 - \theta_1 B - \theta_2 B^2 - \cdots - \theta_q B^q \tag{S1.3}$$

In practice, $d$ is usually 0, 1, or at most 2, where $d = 0$ corresponds to the stationary behavior of the stochastic process. This class of processes are called *autoregressive integrated moving average* (ARIMA) processes of order $(p, d, q)$. The reason for inclusion of the word 'integrated' in ARIMA is as follows. Inverting $\Delta_t = (1-B)^d z_t$ yields $z_t = S^d\Delta_t$, where $S = (1-B)^{-1}$ is the *summation operator* defined by

$$S\Delta_t = \sum_{j=0}^{\infty} \Delta_{t-j} = \Delta_t + \Delta_{t-1} + \Delta_{t-2} + \cdots$$

Thus, a general ARIMA process may be generated by summing or 'integrating' the process $\{\Delta_t\}$ $d$ times.

In what follows $y_t$ denotes the dose metric at fraction $t$, for which the difference in the delivered dose metric to the previous fraction is given by

$$\Delta_t = y_t - y_{t-1} = (1-B)\, y_t. \tag{S1.4}$$

Because the delivered dose metric is allowed to drift over the course of treatment, the dose metric process $\{y_t\}$ is treated as nonstationary but it is assumed that the process of the differences $\{\Delta_t\}$ comes from a stationary process. Then $\{y_t\}$ can be modeled as an ARIMA(0,1,1) process which by equations (S1.1), (S1.2), and (S1.4) is given by:

$$(1-B)\, y_t = (1-\theta_1 B)\, \varepsilon_t \tag{S1.5}$$

where $|\theta_1| < 1$. Inverting, (S1.5) yields:

$$y_t = (1-B)^{-1}(1-\theta_1 B)\, \varepsilon_t, \tag{S1.6}$$

Since $R_t = (1-B)^{-1} y_t$, we find that

$$R_t = (1-B)^{-2}(1-\theta_1 B)\, \varepsilon_t, \tag{S1.7}$$

Hence, with the assumptions we have made for the process $\{y_t\}$ this shows that the nonstationary aggregate $\{R_t\}$ is an ARIMA(0,2,1) process. It can be shown that the process in equation (S1.6) can be written as the weighted average of previous values of the process:

$$y_t = \widehat{y}_{t|t-1}(\alpha) + \varepsilon_t \tag{S1.8}$$
$$\widehat{y}_{t|t-1}(\alpha) = \alpha \sum_{j=1}^{\infty} (1-\alpha)^{j-1} y_{t-j} \tag{S1.9}$$

and $\alpha = 1 - \theta_1$, hence $0 < \alpha < 2$ (cf. Ref. [1]). As discussed in Box et al. [1] $\widehat{y}_{t|t-1}(\alpha)$ is the optimal linear one-step-ahead predictor, also referred to as the level, of the process $\{y_t\}$, i.e. what is known about the future value of this process. Using equation (S1.9) a recursive relationship for this optimal one-step-ahead predictor of $y_t$ is now derived, which shows that $\widehat{y}_{t|t-1}(\alpha)$ is indeed our simple SES model

$$\begin{aligned}
\widehat{y}_{t+1|t}(\alpha) &= \alpha \sum_{j=1}^{\infty} (1-\alpha)^{j-1} y_{t-(j-1)} = \alpha \sum_{k=0}^{\infty} (1-\alpha)^k y_{t-k} \\
\widehat{y}_{t+1|t}(\alpha) &= \alpha y_t + \alpha \sum_{k=1}^{\infty} (1-\alpha)^k y_{t-k} = \alpha y_t + (1-\alpha)\alpha \sum_{k=1}^{\infty} (1-\alpha)^{k-1} y_{t-k} \\
\widehat{y}_{t+1|t}(\alpha) &= \widehat{y}_{t|t-1}(\alpha) + \alpha[y_t - \widehat{y}_{t|t-1}(\alpha)]. \qquad \text{(S1.10)}
\end{aligned}$$

Setting $\ell_t = \widehat{y}_{t+1|t}(\alpha)$ we find

$$\ell_t = \ell_{t-1} + \alpha\,(y_t - \ell_{t-1}) \qquad \text{(S1.11)}$$

This recursive relationship is easily derived from the SES state model shown in equation (3). Thus, the SES model for the per-fraction dose metric is the optimal linear one-step-ahead forecast under an ARIMA(0,1,1) model for the nonstationary dose metric process $\{y_t\}$ and an ARIMA(0,2,1) process for its nonstationary aggregate $\{R_t\}$.

In the traditional SES setting, one restricts $0 < \alpha \leq 1$, for which the SES level is an exponentially weighted moving average with nonnegative weights that sum to one. Although the SES–ARIMA(0,1,1) equivalence implies a broader admissible region $0 < \alpha < 2$ under invertibility, an $\alpha > 1$ yields alternating-sign weights, and therefore does not retain the standard smoothing interpretation; for interpretability in ART one therefore, restricts $\alpha$ to the half open interval $(0, 1]$. Also note that allowing the value $\alpha = 1$, shows that LOCF is one particular case of the SES strategy for adaptive monitoring being discussed in this work (see equation (S1.11) since the level $\ell_t$ for any $t \in \{1, \cdots, N\}$ is then simply equal to the observed value $y_t$.

## S1.2. Derivation of the Remaining-Course Innovation Coefficients, Conditional Mean and Variance

Starting from the state equation at fraction $t$,

$$\begin{aligned}
y_{t+1} &= \ell_t + \varepsilon_{t+1} \\
\ell_{t+1} &= \ell_t + \alpha\varepsilon_{t+1},
\end{aligned}$$

One straightforwardly finds by backwards substitution that

$$y_{t+2} = \ell_t + \alpha\varepsilon_{t+1} + \varepsilon_{t+2},$$

and, more generally for carrying out the indicated backwards substitution $j-1$ times,

$$y_{t+j} = \ell_t + \alpha \sum_{r=1}^{j-1} \varepsilon_{t+r} + \varepsilon_{t+j}.$$

Forming the aggregate of future values by summing from $j = 1$ to $h$ yields

$$R_{t+h} = h\ell_t + \sum_{j=1}^{h} \left( \varepsilon_{t+j} + \alpha \sum_{r=1}^{j-1} \varepsilon_{t+r} \right).$$

To make it clear that the random variable $R_{t+h}$ is based on SES and therefore depends on the level $\ell_t$ and the innovations, we will use the notation $R_{t+h|t}$ hereafter. Note, that a future innovation $\varepsilon_{t+r}$ appears once directly and in each subsequent forecasted observation through the state update. If one writes out the sum over $h$ in lower diagonal matrix form and then sums up each of the resulting columns one finds the total coefficient for a given $\varepsilon_{t+r}$ to be:

$$c_r = 1 + \alpha(h - r), \qquad r = 1, \ldots, h. \qquad \text{(S1.12)}$$

Hence,

$$R_{t+h|t} = h\ell_t + \sum_{r=1}^{h} c_r \varepsilon_{t+r}. \tag{S1.13}$$

The sum in equation (S1.13) can be rewritten as follows

$$\sum_{r=1}^{h} c_r \varepsilon_{t+r} = \sum_{k=0}^{h-1} c_{h-k} \varepsilon_{t+h-k},$$

and hence, equation (S1.13) becomes

$$R_{t+h|t} = h\ell_t + \sum_{k=0}^{h-1} c_{h-k} \varepsilon_{t+h-k}. \tag{S1.14}$$

Using equation (S1.14) one can now easily compute conditional expectation of $R_{t+h}$ as follows;

$$\begin{aligned}
\widehat{R}_{t+h|t} &\equiv E(R_{t+h|t} \mid \mathcal{F}_t) \\
&= E(h\ell_t + \sum_{k=0}^{h-1} c_{h-k} \varepsilon_{t+h-k} \mid \mathcal{F}_t) \\
&= h\ell_t + \sum_{k=0}^{h-1} c_{h-k} E(\varepsilon_{t+h-k} \mid \mathcal{F}_t) \\
&= h\ell_t,
\end{aligned}$$

where we have used following well known fact about the conditional expectation namely, $E(a + bX \mid Y) = a + bE(X \mid Y)$ and that $E(\varepsilon_{t+h-k} \mid \mathcal{F}_t) = 0$ for all $k$ by assumption. For independent innovations as assumed here with common variance $\sigma^2$ one finds for the conditional variance

$$\begin{aligned}
V\left(R_{t+h|t} \mid \mathcal{F}_t\right) &= V\left(h\ell_t + \sum_{k=0}^{h-1} c_{h-k} \varepsilon_{t+h-k} \mid \mathcal{F}_t\right) \\
&= V\left(\sum_{k=0}^{h-1} c_{h-k} \varepsilon_{t+h-k} \mid \mathcal{F}_t\right) \\
&= \sum_{k=0}^{h-1} c_{h-k}^2 V\left(\varepsilon_{t+h-k} \mid \mathcal{F}_t\right) \\
&= \sum_{k=0}^{h-1} c_{h-k}^2 V\left(\varepsilon_{t+h-k}\right) \\
&= \sigma^2 \sum_{k=0}^{h-1} c_{h-k}^2,
\end{aligned}$$

where the following well known facts about the conditional variance of independent random variables have been used in the second, third, and fourth equation, above namely, $V(a + bX \mid Z) = V(bX \mid Z)$, $V(aX + bY \mid Z) = a^2V(X \mid Z) + b^2V(Y \mid Z)$, and $V(X \mid Y) = V(X)$. The last equation follows from the fact that innovations are identically distributed with common variance $\sigma^2$. Using equation (S1.12) yields the following result.

$$\sum_{k=0}^{h} c_{h-k}^2 = \sum_{k=0}^{h-1} (1 + \alpha k)^2 = h + 2\alpha \sum_{k=0}^{h-1} k + \alpha^2 \sum_{k=0}^{h-1} k^2,$$

which after summing over $k$ and simplifying yields the closed-form expression for $\Omega(h, \alpha)$ given in equation (9) in the main text.

# Supplementary Section S2: Empirical Residual Innovation Bootstrap Model

The analytic Gaussian formulation provides a compact, closed-form predictive distribution, but its shape is determined by the assumption of having independent, normally distributed innovations with mean zero and common standard $\sigma^2$. This appendix develops a model-based empirical residual bootstrap in which the observed post-initialization innovations are centered, i.e. have mean zero, and are treated as an empirical distribution for future innovations. These values are propagated through the same SES state dynamics and lead-time coefficients as used in the analytic formulation. The method therefore preserves the underlying forecast structure while allowing the predictive distribution to reflect features of the observed residuals, including skewness, tail asymmetry, and discreteness. Table S2.1 summarizes the principal differences between the two approaches.

**Table S2.1:** Comparison of the analytic and empirical predictive-distribution constructions. FFT: fast Fourier transform; SES: simple exponential smoothing.

| Feature | Analytic Gaussian formulation | Empirical residual-bootstrap formulation |
|---|---|---|
| SES state dynamics | Retained | Retained |
| Innovation distribution | Gaussian with estimated variance | Empirical centered residual distribution |
| Predictive output | Closed-form normal distribution | Discrete predictive probability distribution |
| Computation | Direct formula | Grid construction and FFT convolution |
| Principal advantage | Maximum transparency and minimal computation | Can retain observed skewness, tail asymmetry, or discreteness |
| Principal limitation | Sensitive to non-Gaussian and heteroscedastic innovations | Limited by residual sample size and still assumes future residuals are represented by the empirical distribution of past residuals |

## S2.1. Framework

Denote by

$$\mathcal{E}_t = \{\varepsilon_1, \cdots, \varepsilon_t\}$$

the set of observed one-step-ahead residual innovations from an SES forecasting model. These observed residual innovations are treated as an empirical discrete sample for the future residual innovation distribution. If no weighting of the innovations is assumed, i.e. each innovation has equal probability of occurring, then the empirical $CDF$ can be written as:

$$\widehat{F}_\varepsilon(x) = \frac{1}{t}\sum_{i=1}^{t} H(x - \varepsilon_i),$$

where $H(x)$ denotes the Heaviside step function, which is defined as follows

$$H(x) = \begin{cases} 1 & \text{if } x \geq 0 \\ 0 & \text{if } x < 0. \end{cases}$$

In what follows, we choose to work with the centered residual innovations, i.e residual innovations that are centered around their mean. To this end set

$$\widetilde{\varepsilon}_i = \varepsilon_i - \overline{\varepsilon},$$

where

$$\overline{\varepsilon} = \sum_{i=1}^{t} \varepsilon_i$$

so that future residual innovations are bootstrapped around a mean of zero, rather than around $\bar{\varepsilon}$. The centered $CDF$ then becomes:

$$\widehat{F}_{\tilde{\varepsilon}}(x) = \frac{1}{t}\sum_{i=1}^{t} H(x - \widetilde{\varepsilon}_i).$$

## S2.2. Propagation of Residual Innovations Through the SES State

When doing a multi-time step ahead prediction a future residual innovation can influence more than the single future observation for which it has been sampled. Because the SES state is updated after each future residual innovation, an earlier residual can also affect later forecasts. The cumulative residual contribution to the future total can therefore be written as a weighted residual innovation aggregate:

$$Q_h = \sum_{k=1}^{h} c_j \mathrm{E}_{t+j},$$

where $h$ denotes the number of future time steps, $\mathrm{E}_{t+j}$ denotes the bootstrapped residual random variable at future time step $j$, $c_j$ denotes the SES propagation coefficient for the residual innovation from $j$ to $h$, and $Q_h$ is the random variable representing the total future residual innovation contribution after SES propagation. Note that $Q_h$ is the weighted sum of independent identically distributed random variables.

The coefficients are determined by the SES update equation employed and the selected smoothing and damping parameters. For the update equation, used in this work the $c_j$ are given by $c_j = 1 + \alpha(h - j)$ (see Equation (S1.12) above). The reader is referred to Table 6.6 in Hyndman et al. [2] for the update equations of the most commonly used state space models having homoscedastic innovations.

Observe that the last future residual innovation contributes only directly with coefficient 1, while earlier residual innovations have larger effective coefficients because they also perturb subsequent SES states. We now show how one can estimate the distribution of $Q_h$ using the convolution theorem and Discrete Fourier Transformation (DFT). However, before this can be done, one first needs to define a discrete probability measure for the independent and identically distributed random variables that make up $Q_h$.

## S2.3. Discrete Bootstrap Probability Measures for Future Residual Innovations.

Denote by $(\mathcal{X}_t, \Omega)$ the measurable space one is working with. For each future time step $j$, define the scaled innovation random variable with compact support

$$X_j = c_j \mathrm{E}_{t+j}, \qquad j = 1, \cdots, h.$$

Under the empirical bootstrap described above, the possible realizations for each $X_j$ are

$$x_{i,j} = c_j \widetilde{\varepsilon}_i, \qquad i = 1, \cdots, t$$

If the residual innovations are unweighted, each historical residual innovation contributes with equal probability, namely,

$$p_{i,j} = \frac{1}{t},$$

and one can define the following discrete probability measure for the $X_j$

$$\mu_j(A) = \frac{1}{t}\sum_{i=1}^{t} \delta_{x_{i,j}}(A), \quad j = 1, \cdots, h$$

where $\delta_x$ is the Dirac measure, which is defined as follows. For a fixed point $x \in \mathcal{X}_t$, the Dirac measure is defined for any set $A \in \Omega$ as

$$\delta_x(A) = \begin{cases} 1 & \text{if } x \in A \\ 0 & \text{if } x \notin A. \end{cases}$$

Since DFT operates on arrays, the values of the scaled innovation random variables $X_j$ need to be placed on a common grid. Let $q < 1$ be the quantization step width, then one can define a common grid by rounding to the closest integer

$$k_{i,j} = \lfloor q x_{i,j} \rceil, \qquad i = 1, \cdots, t$$

where $\lfloor \cdot \rceil$ denotes the operation to round to the closest integer. The step specific probability measure associated with this grid definition is then given by

$$\mu_j(A) = \frac{1}{t} \sum_{i=1}^{t} \delta_{k_{i,j}}(A), \qquad j = 1, \cdots, h.$$

## S2.4. Why FFT Avoids Bootstrap Combinatorics

A direct residual-bootstrap enumeration would consider all possible paths

$$(\mathrm{E}_{t+1},\ \mathrm{E}_{t+2}, \cdots, \mathrm{E}_{t+h}\ ).$$

With $t$ historical residual innovations and $h$ future time steps, a complete enumeration of paths requires enumeration of $t^h$ distinct paths. Each path produces one specific realization of $Q_h$, namely,

$$q_h^{(p)} = \sum_{j=1}^{h} c_{h-j} \widetilde{\varepsilon}_{p_j}, \qquad \{p_1, \cdots, p_h\} \in \{1,\ \cdots, t\}^h$$

where $\{p_1, \cdots, p_h\}$ denotes one of the $t^h$ possible permutations with replacement generated from the set $\{1,\ \cdots, t\}$. This enumeration is combinatorial. Complete enumeration quickly becomes impractical as $h$ and $t$ grow beyond small numbers: for 10 past time steps and a prediction horizon of 5, one would already have to enumerate $10^5$ possible paths. Monte Carlo bootstrap avoids complete enumeration by using random sampling of paths; however, this introduces random sampling error, especially in the predictive tails of the estimated bootstrap distribution [3].

The DFT approach instead computes the discrete bootstrap probability measure associated with $Q_h$ by convolution of the probability measures associated with each of the $X_j,\ j = 1, \cdots, h$.

## S2.5. Computation of $\mu_{Q_h}$ using Convolution and the Discrete Fourier Transformation

Since $Q_h$ is the sum of independent and identically distributed random variables assumed to have a common compact support,

$$Q_h = \sum_{j=1}^{h} X_j,$$

it can be bootstrapped using convolution and DFT [4]. The probability measure for $Q_h$ can be found by convolution of the step specific probability measures

$$\mu_{Q_h} = \mu_1 * \mu_2 * \cdots * \mu_h$$

Applying DFT recursively to this expression and using the convolution theorem yields the following result.

$$\mathcal{F}\left(\mu_{Q_h}\right) = \prod_{j=1}^{h} \mathcal{F}\left(\mu_j\right),$$

and inversion yields

$$\mu_{Q_h} = \mathcal{F}^{-1}\left(\prod_{j=1}^{h} \mathcal{F}\left(\mu_j\right)\right)$$

Here DFT is used neither to estimate a residual spectrum nor to infer periodic structure, but to convert the operation of convolution to multiplication in the frequency domain and to transform the resulting product back to the time domain to compute the bootstrap probability measure for $Q_h$.

## S2.6. From the Residual Sum *CDF* to the Predictive *CDF*

Once the probability measure for $Q_h$ has been computed, it can be shifted and scaled to obtain the predictive discrete probability measure for the final quantity of interest. If $D_t$ denotes the deterministic forecast aggregate and $N$ the total number of fractions, then

$$\overline{Y} = g(Q_h) = \frac{D_t + Q_h}{N}.$$

Since $g$ is one to one, the $CDF$ for $\overline{Y}$ can be determined from the $CDF$ of $Q_h$as follows:

$$F_{\overline{Y}}(y) = P\left(\overline{Y} \leq y\right) = P\left(Q_h \leq g^{-1}(y)\right) = P\left(Q_h \leq Ny - D_t\right) = F_{Q_h}\left(Ny - D_t\right).$$

Thus, the DFT convolved discrete residual bootstrap probability measure directly generates a predictive discrete probability measure for the final dose metric. The method is nonparametric with respect to the distribution of the residual innovations because the shape of the discrete $\mu_{Q_h}$is inherited from the empirical discrete sample of the residual innovations rather than imposed by a Gaussian probability measure.

### S2.6.1 Predictive Performance Comparison to the Gaussian distribution

Predictive performance of the centered empirical residual-bootstrap and analytic Gaussian distributions was compared using the continuous ranked probability score (CRPS), which evaluates the complete predictive distribution against the realized value, with lower scores indicating better distributional performance [5]. For each sequence and forecast horizon, the relative paired difference was calculated as $\left(CRPS_{bootstrap} - CRPS_{Gaussian}\right) / CRPS_{Gaussian}$, so negative values favor the empirical FFT distribution and positive values favor the Gaussian distribution.

For visualization, cohort identity was displayed consistently across the supplementary comparisons: the total cohort is shown in dark gray, head and neck in blue, and parotid in orange. Reference lines were neutral, so the sign of a paired difference was conveyed by its position relative to zero rather than by color.

At $\alpha = 0.5$, the median relative differences were negative in 10 of the 12 cohort–horizon comparisons (Figure S2.1). For the total cohort, the medians were +0.02%, −0.72%, −1.41%, and −1.98% at $h = 5$, 10, 15, and 20, respectively. The corresponding medians for the head-and-neck cohort were +0.19%, −0.99%, −1.87%, and −2.45%. While medians for the Parotid cohort were −0.68%, −1.53%, −1.78%, and −0.73%.

The median differences were therefore close to zero at $h = 5$ for the total and head-and-neck cohorts and modestly favored the empirical FFT distribution at longer horizons. The parotid medians were negative at all four horizons. The distributions nevertheless extended across zero in the displayed comparisons, indicating substantial overlap in per-series performance. The relative differences are small indicating the Gaussian approximation is suitable for LT-SES.

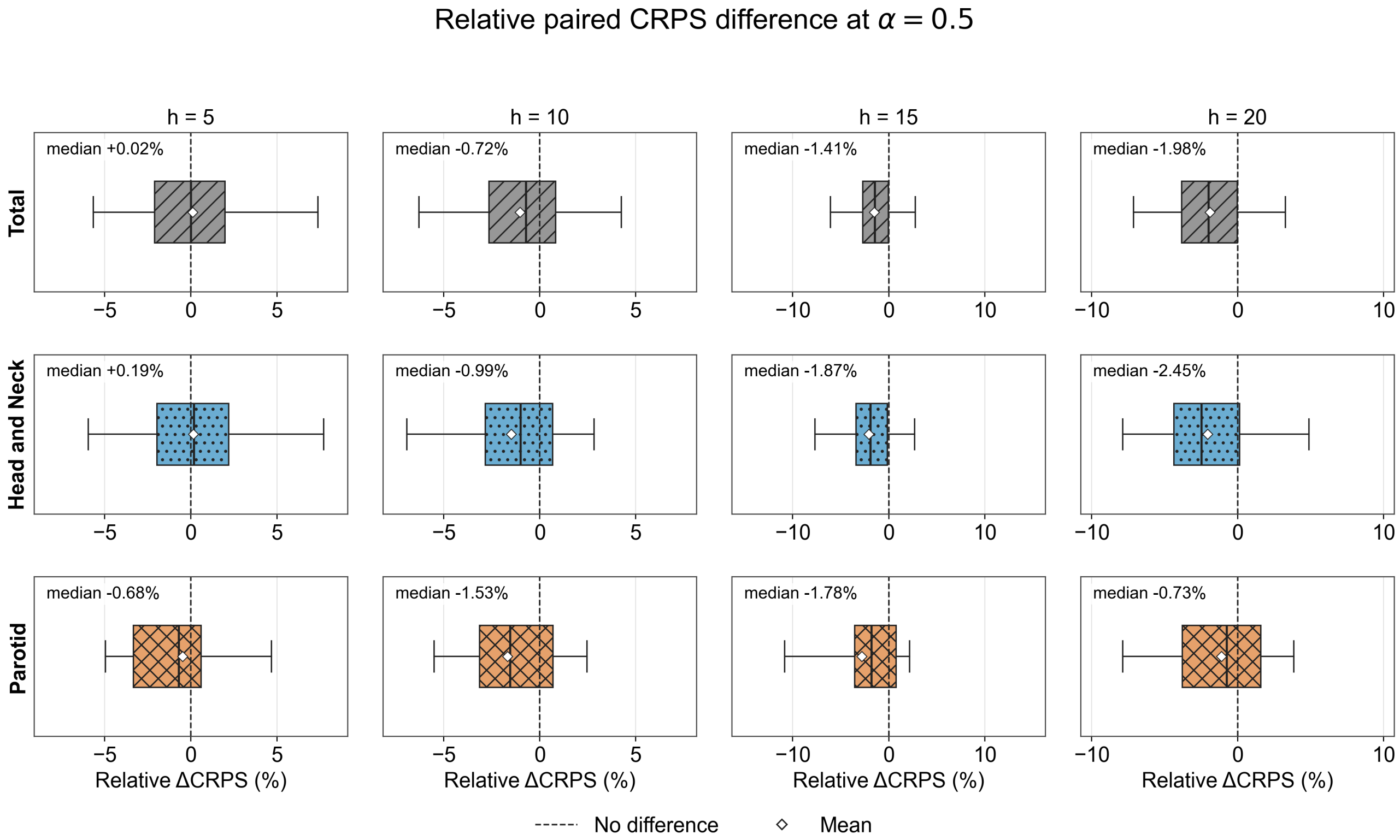

**Figure S2.1:** Relative paired continuous ranked probability score (CRPS) differences at $\alpha = 0.5$ for the total cohort ($n = 340$; top), head-and-neck cohort ($n = 181$; middle), and parotid cohort ($n = 44$; bottom), comparing the analytic Gaussian and centered empirical FFT-bootstrap predictive distributions. Positive values favor the Gaussian distribution and negative values favor the empirical FFT distribution. Boxes show interquartile ranges, whiskers show the 5th–95th percentiles, and diamonds show arithmetic means.

# Supplementary Section S3: Descriptive Predictive Comparison of SES and Alternative Forecasting Models

## Methods

SES using three fixed $\alpha$-values (1.00, 0.50, and 0.25) was compared with an exchangeable Bayesian reference, series specific SES, Holt's linear method, and damped Holt's method. Forecasts used $h = 5$, 10, 15, or 20 fractions remaining and only data through $n = N - h$. The target was the realized final arithmetic mean,

$$\bar{y}_N = \frac{1}{N}\sum_{t=1}^{N} y_t. \tag{S3.1}$$

Each organ-at-risk structure time series was analyzed separately. Results are shown for the full cohort and the nested head-and-neck and parotid groups.

The Bayesian model assumed exchangeable normal observations with unknown mean and variance and prior $\pi(\mu, \sigma^2) \propto 1/\sigma^2$. Its predictive distribution for the final mean was

$$\overline{Y}_N \mid y_{1:n} \sim t_{n-1}\left(\bar{y}_n,\ s_n\sqrt{\frac{h}{n(n+h)}}\right), \tag{S3.2}$$

where $s_n$ is the sample standard deviation of the observed prefix and the second argument denotes the Student-$t$ scale. Because the Bayesian forecast depends only on the prefix mean and standard deviation, reordering observations does not change it. The exponential-smoothing models retain time order: SES tracks a local level, Holt's linear method an undamped trend, and damped Holt's method a damped trend. These correspond to ETS(A,N,N), ETS(A,A,N), and ETS(A,A$_d$,N), respectively [6–8].

Visual encodings were held consistent across the supplementary comparisons. Cohort summaries use dark gray for the full cohort, blue for head and neck, and vermillion for parotid, with distinct line, marker, or hatch patterns where applicable. Model colors are held constant across panels. In the HN3 and HN15 examples, SES with $\alpha = 0.25$ is shown as a blue solid curve and the Bayesian reference as a dark-gray dashed curve, with matching shaded prediction intervals and neutral threshold lines. Printed labels and distinct line or marker patterns provide additional identification.

**Table S3.1:** Fitted quantities and the corresponding K used in the exponential-smoothing innovation-variance estimate.

| Model | Fitted from the observed prefix | Plain-language role | K |
|---|---|---|---|
| Bayesian reference | $\mu$ and $\sigma^2$ integrated over their posterior | Common mean and between-fraction spread; all observations have equal weight | N/A |
| SES, $\alpha = 1.00$ | Initial level $\ell_0$ | Starting level; the current level then follows the latest observation | 1 |
| SES, $\alpha = 0.50$ | Initial level $\ell_0$ | Starting level with moderate response to new observations | 1 |
| SES, $\alpha = 0.25$ | Initial level $\ell_0$ | Starting level with a smoother, slower response | 1 |
| Series-specific SES | $\ell_0$ and $\alpha$ | Starting level and fitted responsiveness | 2 |
| Holt's linear method | $\ell_0$, initial trend $b_0$, $\alpha$, and $\beta$ | Level plus an undamped trend; $\beta$ controls trend updating | 4 |
| Damped Holt's method | $\ell_0$, $b_0$, $\alpha, \beta$, and damping $\phi$ | Level plus a trend whose future effect decreases when $\phi < 1$ | 5 |

*Note.* K is the number of parameters fitted before estimating the innovation variance; the innovation variance itself is not included. The Bayesian model uses its Student-$t$ posterior predictive distribution and does not use this correction.

At each forecast origin, all free quantities were estimated from the available prefix by minimizing the sum of squared one-step ahead innovations (SSE). For fixed-$\alpha$ SES, only $\ell_0$ was fitted. For series-specific SES, $\ell_0$ was re-estimated for each candidate $\alpha$ before selecting the pair with the lowest error. Holt's linear and damped Holt methods were fitted in the same manner. The plug-in innovation variance was

$$\widehat{\sigma}_n^2 = \frac{\mathrm{SSE}_n}{n - \mathrm{K}}. \tag{S3.3}$$

The fitted state and covariance were propagated over the remaining $h$ fractions, combined with the observed prefix sum, and divided by $N$ to obtain the final-mean predictive distribution. Predictive performance was characterized using CRPS, for which lower values are better [5]. CRPS values were normalized after forecasting by the realized full-course mean.

A separate cohort profile characterized one-step SES error. Each complete sequence was normalized by its full-course mean, $\ell_0$ was re-estimated for every candidate $\alpha$, and squared innovations were pooled within each cohort. This profile was descriptive; the predictive comparison used the same fixed value, $\alpha = 0.25$, for all cohorts.

## Results

The profiled one-step mean squared error (MSE) attained its minimum at $\alpha = 0.22$ for the full cohort, 0.25 for the head and neck cohort, and 0.27 for the parotid cohort (Figure S3.1). As can be seen from Figure S3.1 the curves were shallow near their minima, and the common value $\alpha = 0.25$ was close to all three. These values describe this particular dataset and are not clinical prescriptions.

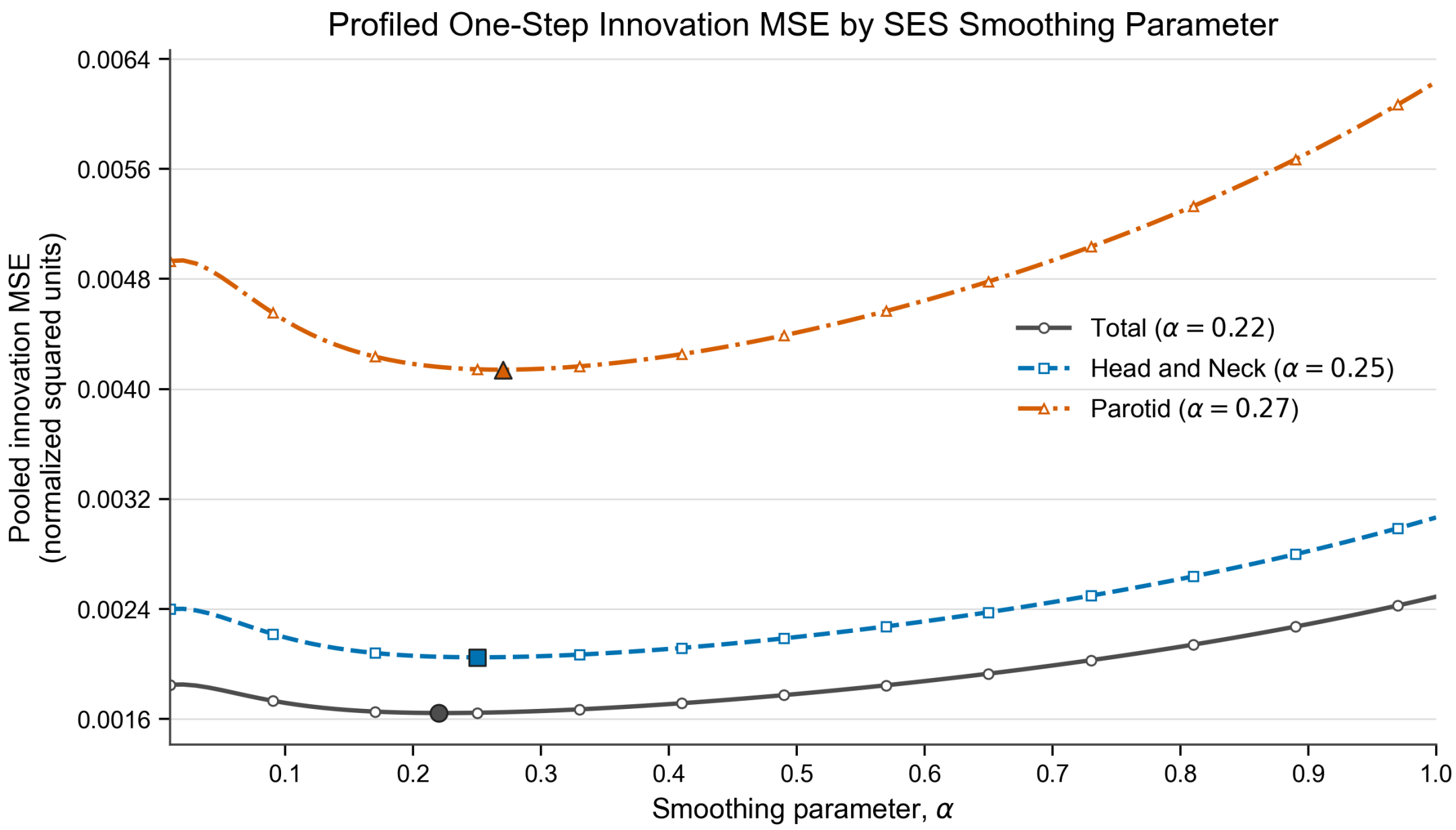

**Figure S3.1:** Profiled one-step innovation MSE by SES smoothing parameter. The initial level was re-estimated for every candidate $\alpha$. The full cohort is shown in dark gray with circles and a solid line, the head and neck cohort in blue with squares and a dashed line, and the parotid cohort in vermillion with triangles and a dash-dot line. The later predictive comparison used fixed $\alpha = 0.25$ in all cohorts.

Fixed SES with $\alpha = 0.25$ had the lowest arithmetic-mean CRPS in all 12 cohort–horizon panels (Figure S3.2). The Bayesian model and SES with $\alpha = 0.50$ were generally the closest alternatives, with substantial overlap among distributions. Series-specific fitting of $\alpha$ did not provide a consistent improvement. Holt's linear method, which includes an undamped trend, produced the largest long-horizon errors; damped Holt reduced, but did not remove, this behavior by damping the trend.

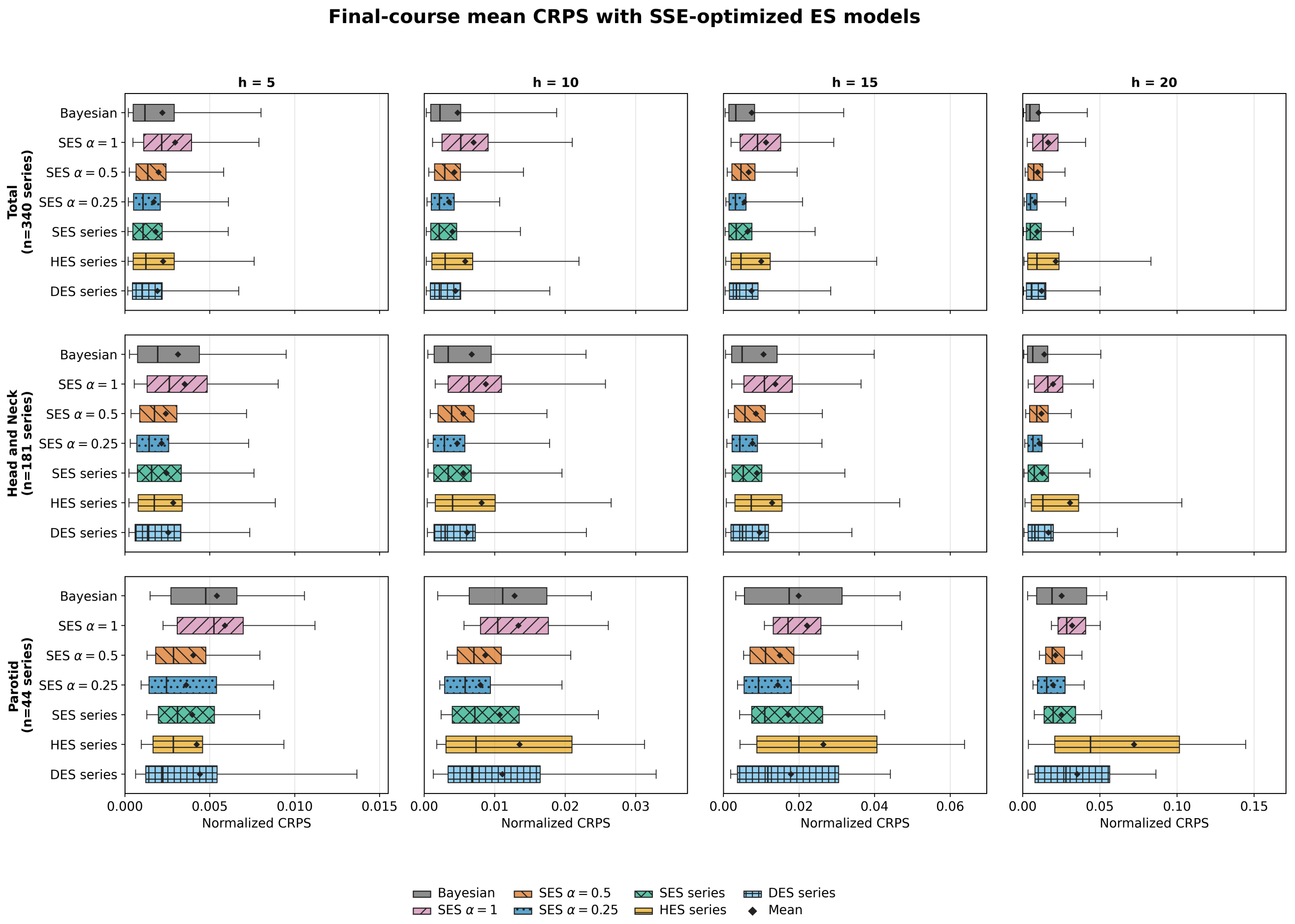

**Figure S3.2:** Per-series normalized CRPS at horizons of 5, 10, 15, and 20 remaining fractions. Plot labels HES and DES correspond to Holt's linear method and damped Holt's method, respectively. Model colors are held constant across panels: Bayesian is dark gray; SES with $\alpha$ = 1.0, 0.50, and 0.25 is reddish purple, vermillion, and blue, respectively; series-specific SES is bluish green; HES is orange; and DES is sky blue. Boxes retain neutral outlines and printed model labels. Boxes show interquartile ranges, center lines show medians, whiskers show the 5th and 95th percentiles, and diamonds show arithmetic means. Lower CRPS is better.

The within-series comparison showed the same general pattern (Figure S3.3). Differences were calculated as Bayesian CRPS minus SES CRPS, so positive values favor SES with $\alpha = 0.25$. Medians and arithmetic means were positive in every panel, but the distributions also extended below zero. Thus, SES performed better for the typical series, but not for every series.

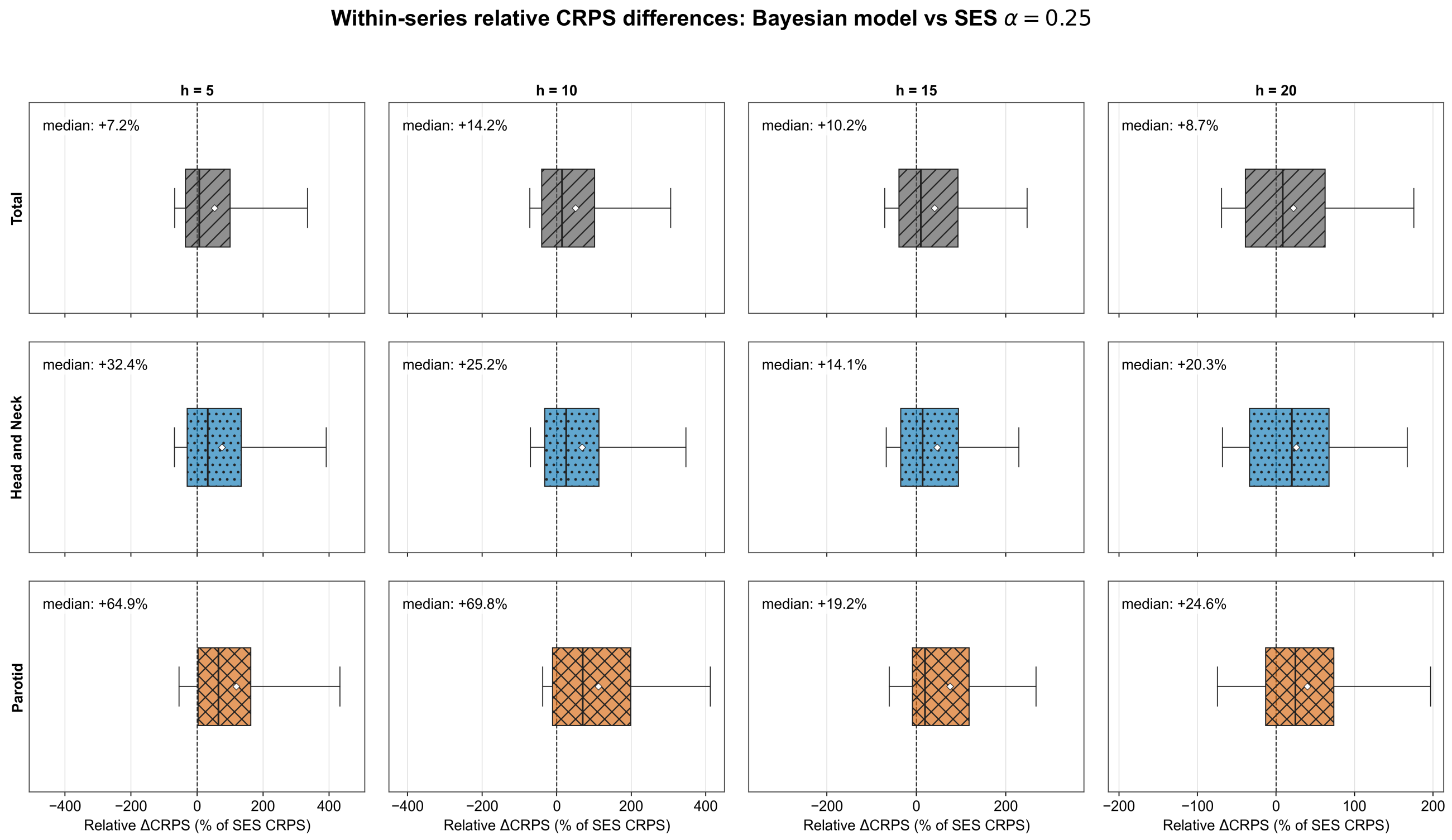

**Figure S3.3:** Within-series normalized CRPS difference between the Bayesian model and SES with fixed $\alpha = 0.25$. The full, head-and-neck, and parotid cohorts use the common dark-gray, blue, and vermillion mapping, respectively, with a neutral zero-reference line. Positive values favor SES and negative values favor the Bayesian model by position relative to zero. These are descriptive structure-level comparisons, not patient-level confidence intervals.

## Parotid Examples

The examples illustrate why time order may have an impact on patient management during ART. The Bayesian model gives equal weight to all observed fractions, whereas SES places the highest weight to the most recent observations. In HN3, which has marked day-to-day variation, SES responds sooner to recent low- and high-dose clusters, while the Bayesian forecast is more stable (see Figure S3.4). This responsiveness changes the projected final mean and threshold probabilities earlier, but it can also reflect short-lived variation.

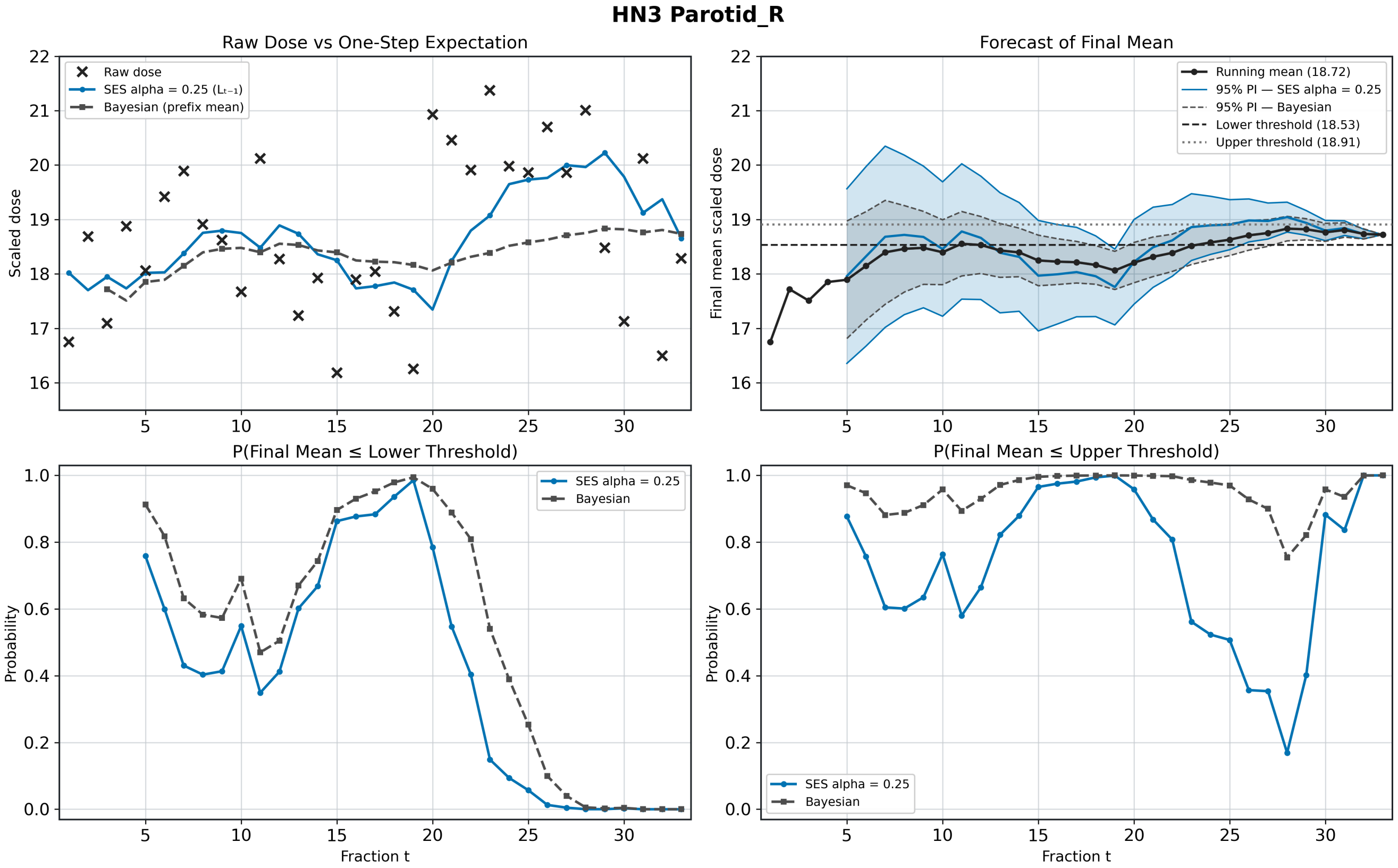

**Figure S3.4:** HN3 right-parotid example. SES with $\alpha = 0.25$ responds sooner to recent clusters of observations, whereas the exchangeable Bayesian model changes through the cumulative prefix mean and variance. SES is shown as a blue solid curve and the Bayesian reference as a dark-gray dashed curve, with matching blue and gray prediction-interval fills; threshold lines are neutral and distinguished by dash pattern. Thresholds are illustrative and are not clinical action levels.

In example HN15, a sustained increase begins near the middle of the course (see Figure S3.5). SES moves toward the new local level sooner. Its lower-threshold probability falls sharply around fractions 20 – 22, and its upper-threshold probability temporarily decreases, while the Bayesian upper-threshold probability remains near one. The forecasts converge only after more post-change observations accumulate.

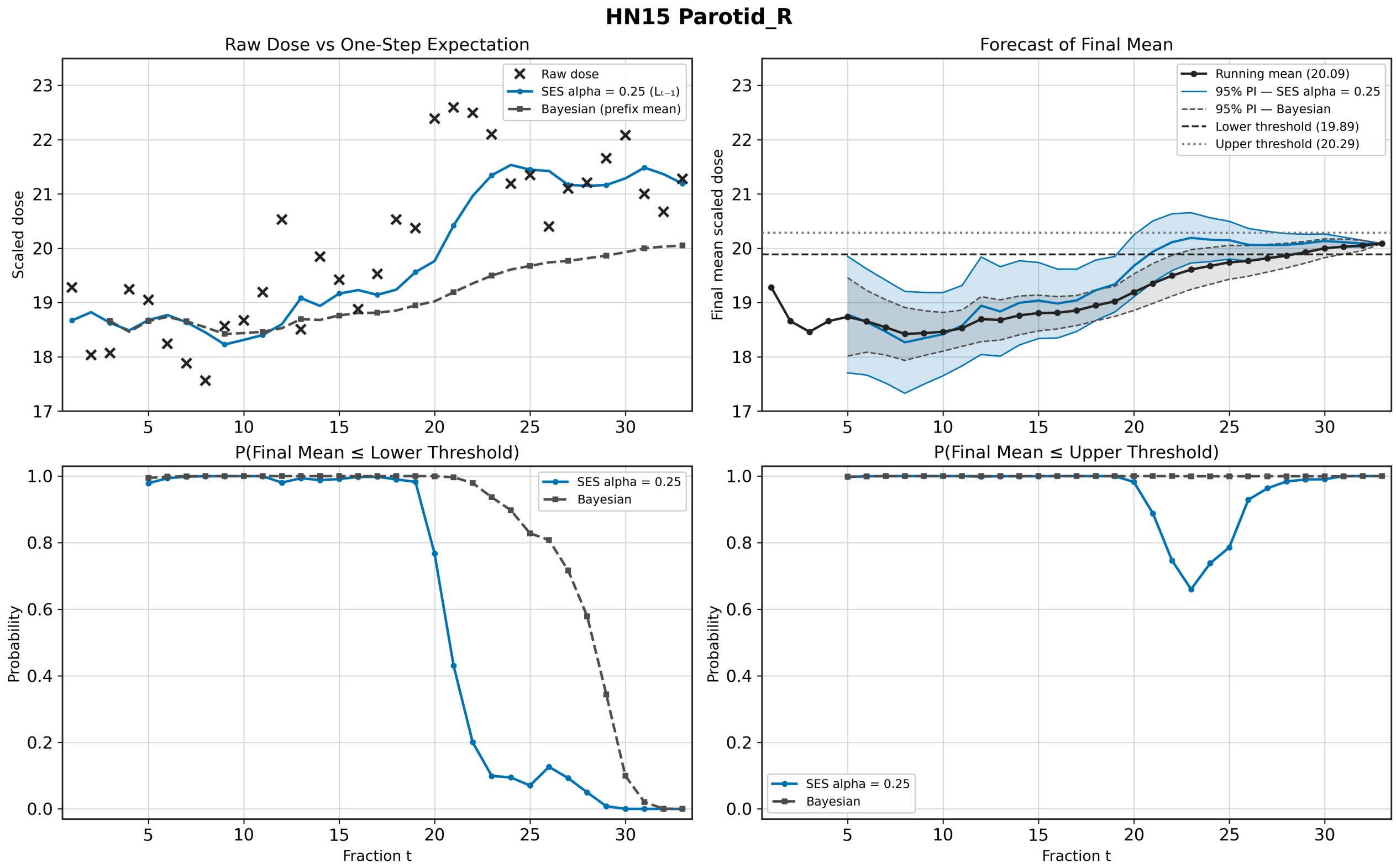

**Figure S3.5:** HN15 right-parotid example with a sustained midcourse increase. SES with $\alpha = 0.25$ adapts sooner to the new local level; the Bayesian prefix mean changes more slowly because observation order does not affect the model. SES is shown as a blue solid curve and the Bayesian reference as a dark-gray dashed curve, with matching blue and gray prediction-interval fills and neutral threshold lines.

## Interpretation and Limitations

Overall, the results support the use of a relatively simple local-level model for these sparse fraction-level time series. These findings are specific to the measured dataset evaluated here and should not be interpreted as establishing a universally optimal model or value of $\alpha$. The fixed value of $\alpha = 0.25$ was a useful retrospective reference; clinical selection of $\alpha$ should still reflect the desired balance between responsiveness and stability. This selection configures the monitoring response and does not calibrate the resulting threshold probabilities. For ranking-only applications evaluation should focus on ordering and comparability; absolute probability interpretation requires a separate probability-calibration assessment. The cohorts were nested, several structures came from the same patient, and only one forecast origin was tested for each time series and prediction horizon. The results therefore describe time series-level performance rather than patient-level inference. Residual autocorrelation, changing variance, nonlinear drift, and abrupt regime changes were not explicitly modeled.

# Supplementary Section S4: Practical Considerations and Institutional Experience

The preceding sections describe the technical basis and exploratory evaluation of LT-SES. This section provides a different perspective. It summarizes the practical experience that motivated the framework and the choices that were made by us while developing it for adaptive radiotherapy monitoring. These observations reflect our institutional experience and should not be interpreted as validated clinical recommendations.

## S4.1. Why We Pursued LT-SES

Our work began by exploring alternatives to the last-observation-carried-forward (LOCF) approach used by the commercial software available in our clinical workflow. LOCF provided a simple projection of the most recent fraction-level dose metric across the remaining treatment course, but we wanted an approach that could incorporate information from the preceding realized values in the fraction time series rather than relying entirely on the latest observation. This led us to explore simple exponential smoothing and subsequently to use SES for process control of patient treatments, as described in our earlier work [9]

As we gained experience with SES, we wanted to provide more information than either process-control output or a simple scalar difference between an observed or projected dose metric and its clinical goal. In particular, we wanted the monitoring result to reflect both the variation observed during treatment and the amount of treatment remaining. This led to the lead-time formulation developed in the present work, in which the projected final-course metric is accompanied by an estimate of its predictive uncertainty.

The commercial adaptive-radiotherapy software provided useful fraction-level dosimetric information, but its reporting tools did not provide the longitudinal screening and prioritization we needed. Reviewing every patient and every dose metric individually each day was not practical. Therefore, we developed an institutional dashboard that summarized ongoing treatments using information generated by the commercial software.

The dashboard allowed us to filter dose metrics according to their LT-SES probability. One useful filter identified dose metrics with probabilities between 0.50 and 0.95. In practice, limiting the initial review to dose metrics within this range reduced the number of patients requiring attention to a manageable level. Any dose metric that entered this range at any time during the treatment course was designated as a targeted dose metric and remained under ongoing evaluation for the remainder of treatment. Patients could then be ranked according to their most concerning targeted dose metrics, turning a large amount of daily information into a focused worklist.

## S4.2. A Practical Choice of Dose Metrics

Our decision to use fraction-averaged scaled dose metrics (Tier 2) was largely practical (see Section 2.1 for the definition of Tier 1, Tier 2, & Tier 3 dose metrics). This approach allowed manual correction of contours generated by deformable propagation without requiring refinement of the underlying deformation map and opened the possibility of using AI autocontouring on synthetic image-of-the-day CTs. Our quality assurance of deformable dose accumulation, informed by DVH-overlay methods for evaluating deformable-registration-based dose accumulation [10], showed errors comparable in magnitude to the uncertainties associated with arithmetic summation of dose metrics and DVHs. We therefore favored a dose-of-the-day approach, using published Tier 1 clinical goals as reference thresholds while following the Tier 2 metrics longitudinally over the course of treatment.

## S4.3. Choosing Alpha for Routine Monitoring

In practice, we found greater utility in choosing $\alpha$ according to the desired response of the monitoring signal than in using an $\alpha$ optimized separately for each individual time series. The purpose of $\alpha$ was therefore to configure how quickly the monitoring system responded to new observations, balancing sensitivity to sustained change against stability in the presence of short-term variation.

We used cohort-level one-step-ahead prediction error to help establish the lower end of the useful range for $\alpha$. Across the cohorts we examined, this suggested $\alpha$-values of approximately 0.20–0.30. We regarded this range as a practical lower bound for selecting $\alpha$, rather than as an optimal setting for clinical use.

Values above this lower range could then be selected prospectively according to the desired signal response. This approach proved more useful operationally than time series-level optimization because it allowed the same temporal-response setting to be applied consistently across patients and dose metrics, including dose-volume metrics, while preserving $\alpha$ as an explicit and clinically interpretable configuration of the monitoring system.

## S4.4. From Triggers to Prioritization

We initially expected that LT-SES probabilities could be converted into fixed trigger values for adaptive review. In practice, establishing useful probability triggers proved difficult. The significance of a probability depended on the dose metric, the number of treatment fractions remaining, the potential benefit of adaptation, and the resources available to act. A single probability value therefore did not naturally translate into a universal clinical action.

Prioritization, on the other hand, proved to be a more natural use of the information. Once dose metrics had been identified for ongoing evaluation, patients could be ordered according to their most concerning targeted dose metrics and reviewed from the top of the list downward. This allowed clinical attention to be directed where it was most needed without requiring a probability value to function as an automatic trigger.

Importantly, evaluation did not have to wait until a conventional trigger had been reached. Patients whose treatment trajectories appeared increasingly concerning could be queued for imaging review, contour review, dose evaluation, or possible adaptation while sufficient treatment fractions remained for an intervention to be useful.

This proactive approach fit our available resources better than either reviewing every patient intensively or waiting for a late-course trigger. Adaptive evaluation, replanning, plan review, and quality assurance require substantial clinical effort. Prioritization allowed that effort to be directed toward patients most likely to warrant closer examination and allowed those examinations to be scheduled efficiently.

## S4.5. Uncertainty and the Value of a Simple Monitoring Layer

Both simple exponential smoothing and Tier 2 dose metric monitoring involve approximations. In our experience, however, these uncertainties were generally small compared with those already present in the measurement layer. The reliability of the monitored dose metric depends on daily image quality, dose-calculation accuracy, the accuracy of image registration and deformation when used, and the quality of automatic contouring. Errors or inconsistencies in these steps are carried directly into the dose metric subsequently analyzed by LT-SES.

This relative importance was not formally quantified in the present study and should therefore be understood as an observation from our clinical experience. Nevertheless, it influenced our decision to favor a relatively simple forecasting model. A more advanced statistical model cannot compensate for inaccurate or inconsistent measurements supplied to it.

As the measurement layer improves, this balance may change. Better imaging, dose calculation, automatic contouring, registration, and dose accumulation may justify the use of more advanced forecasting models. With sufficiently reliable measurements and appropriate prospective validation, predictive probabilities may eventually be calibrated to clinically meaningful actions or intervention levels.

Our initial goal was more modest. We accepted the limitations of a simple LT-SES model and Tier 2 monitoring because the dashboard still provided an additional layer of safety monitoring for patients treated in our clinic. It systematically screened ongoing treatments, identified metrics for continued attention, and helped direct limited adaptive resources toward patients who warranted closer examination. Most importantly, this could occur proactively, while there was still time to investigate the treatment and consider adaptation if appropriate.

We consider this experience an early step in developing practical, uncertainty-aware monitoring for adaptive radiotherapy. The methods described here will require further refinement and validation as measurement and adaptive workflows continue to improve. We hope this work encourages further interest in these questions within the medical physics community and stimulates additional development of practical approaches to longitudinal treatment monitoring and adaptive decision support.